\documentclass[a4paper,11pt]{article}
\makeatletter
\g@addto@macro\bfseries{\boldmath}
\makeatother
\usepackage[english]{babel}
\usepackage{jheppub}
\pdfoutput=1
\usepackage[T1]{fontenc} % if needed
\usepackage{bbold}
\usepackage{amsmath}
\usepackage{empheq}
\usepackage{booktabs}
\usepackage{color}
\usepackage[utf8]{inputenc}
\usepackage{xspace}
\usepackage{scalerel}
\usepackage[most]{tcolorbox}
\usepackage{multirow}
\usepackage{scalefnt}
\usepackage{bold-extra}
\usepackage{afterpage}
\usepackage{ulem}

\allowdisplaybreaks

\usepackage{xspace}
\usepackage[ddmmyyyy,hhmmss]{datetime}

\newcommand\F{${\rm F}$}
\newcommand\FJ{${\rm FJ}$}
\newcommand\FJJ{${\rm FJJ}$}
\newcommand\PhiB{\Phi_{\scriptscriptstyle \rm F}}

\newcommand\PhiBJ{\Phi_{\scriptscriptstyle \rm FJ}}

\newcommand{\as}{\alpha_s}

\newcommand{\pt}{{p_{\text{\scalefont{0.77}T}}}}

\newcommand{\ptrad}{{p_{\text{\scalefont{0.77}T,rad}}}}

\newcommand{\ptj}{{p_{\text{\scalefont{0.77}T,$j$}}}}
\newcommand{\ptjone}{{p_{\text{\scalefont{0.77}T,$j_1$}}}}
\newcommand{\yjone}{{y_{\text{\scalefont{0.77}$j_1$}}}}
\newcommand{\ptl}{{p_{\text{\scalefont{0.77}T,$\ell$}}}}
\newcommand{\ptlp}{{p_{\text{\scalefont{0.77}T,$\ell^+$}}}}
\newcommand{\mll}{{m_{\text{\scalefont{0.77}$\ell\ell$}}}}
\newcommand{\ptll}{{p_{\text{\scalefont{0.77}T,$\ell\ell$}}}}

\newcommand{\etal}{{\eta_{\text{\scalefont{0.77}$\ell$}}}}

\newcommand{\ylp}{{y_{\text{\scalefont{0.77}$\ell^+$}}}}
\newcommand{\yll}{{y_{\text{\scalefont{0.77}$\ell\ell$}}}}

\newcommand{\muF}{{\mu_{\text{\scalefont{0.77}F}}}}
\newcommand{\muR}{{\mu_{\text{\scalefont{0.77}R}}}}
\newcommand{\KF}{K_{\text{\scalefont{0.77}F}}}
\newcommand{\KR}{K_{\text{\scalefont{0.77}R}}}

\newcommand{\rad}{\mathrm{rad}}
\DeclareMathOperator{\sgn}{sgn}

\newcommand{\noun}[1]{{\scshape #1}}

\newcommand{\POWHEG}{\noun{POWHEG}}

\newcommand{\minlo}{{\noun{MiNLO$^{\prime}$}}}
\newcommand{\minnlo}{{\noun{MiNNLO$_{\rm PS}$}}}
\newcommand{\Matrix}{{\noun{Matrix}}}
\newcommand{\PYTHIA}[1]{\noun{Pythia{#1}}}

\newcommand{\abarmu}[1]{\frac{\as(#1)}{2\pi}}

\newcommand{\citere}[1]{ref.\,\cite{#1}}
\newcommand{\citeres}[1]{refs.\,\cite{#1}}
\newcommand{\eqn}[1]{eq.\,(\ref{#1})}
\newcommand{\neqn}[1]{eqs.\,(\ref{#1})}
\newcommand{\fig}[1]{figure\,\ref{#1}}

\newcommand{\Figs}[1]{Figures\,\ref{#1}}

\newcommand{\Tab}[1]{Table\,\ref{#1}}
\newcommand{\sct}[1]{section\,\ref{#1}}

\newcommand{\app}[1]{appendix\,\ref{#1}}

\newcommand{\LambdaPWG}{\Lambda_{\rm pwg}}

\newcommand{\Eq}[1]{Eq.\,\eqref{eq:#1}}
\newcommand{\eq}[1]{eq.\,\eqref{eq:#1}}
\newcommand{\eqs}[2]{eqs.\,\eqref{eq:#1} and \eqref{eq:#2}}

\renewcommand{\sec}[1]{section\,\ref{sec:#1}}

\newcommand{\bT}{\mathbf{T}}

\newcommand{\nn}{\nonumber}
\newcommand{\df}{\mathrm{d}}
\newcommand{\img}{\mathrm{i}}
\newcommand{\eps}{\epsilon}
\newcommand{\Ecm}{E_{\rm cm}}

\newcommand{\cL}{\mathcal{L}}
\newcommand{\cM}{\mathcal{M}}

\newcommand{\cO}{\mathcal{O}}
\newcommand{\cS}{\mathcal{S}}
\newcommand{\Tau}{\mathcal{T}}
\newcommand{\TauCut}{\mathcal{T}^{\rm cut}}
\newcommand{\bn}{{\bar n}}
\newcommand{\sing}{\mathrm{sing}}

\newcommand{\GammaC}{\Gamma_\mathrm{C}}

\usepackage{xcolor}
\newcommand{\mathd}{\mathrm{d}}
\newcommand{\tmop}[1]{\ensuremath{\operatorname{#1}}}

\newtcolorbox{empheqboxed}{colback=white!35,
 colframe=black,
 width=\textwidth,
 sharpish corners,
 top=-2mm, % default value 2mm
 bottom=0pt
}

\title{Jettiness formulation of the {M{\scalefont{0.77}I}NNLO$_{\text{PS}}$} method}

\author[a]{Markus Ebert,}
\author[b]{Luca Rottoli,}
\author[a]{Marius Wiesemann,}
\author[a,c]{Giulia Zanderighi}
\author[a,d]{\\and Silvia Zanoli}

\emailAdd{ebert@mpp.mpg.de}
\emailAdd{luca.rottoli@uzh.ch}
\emailAdd{marius.wiesemann@mpp.mpg.de}
\emailAdd{zanderi@mpp.mpg.de}
\emailAdd{silvia.zanoli@physics.ox.ac.uk}

\affiliation[a]{Max-Planck-Institut f\"ur Physik, Boltzmannstr. 8  
D-85748 Garching, Germany}

\affiliation[b]{University of Zurich, Winterthurerstrasse  190, 8057 Zurich, Switzerland}
\affiliation[c]{Physik-Department, Technische Universit\"at M\"unchen, James-Franck-Strasse 1, 85748 Garching,
  Germany}
\affiliation[d]{Rudolf Peierls Centre for Theoretical Physics, Clarendon Laboratory, Parks Road, University of
Oxford, Oxford OX1 3PU, UK}

\abstract{We present a new formulation of the \minnlo{} method to match 
NNLO QCD calculations with parton showers by using jettiness
as a resummation variable. The full derivation for colour-singlet processes is presented using $0$-jettiness  starting from the NNLL$^\prime$ resummation formula. We show phenomenological results 
for Drell-Yan and Higgs-boson production at the LHC and compare our predictions to ATLAS and CMS data. 
Differences to the original \minnlo{} formulation using the transverse momentum
of the colour singlet as resummation variable are discussed. We further present a comparison 
of \minnlo{} predictions with {\sc Geneva}. 
Finally, we extend the formulation of the \minnlo{} method to $1$-jettiness which is applicable to processes with a colour singlet plus one jet in the final state.}


\preprint{\\MPP-2024-14 \\OUTP-24-01P \\ZU-TH 06/24}

\begin{document}

\maketitle

%============================================================================
\section{Introduction}
\label{sec:intro}

Accurate simulations have become the theoretical cornerstone of many
physics analyses at the Large Hadron Collider (LHC) today.  They are
mandatory not only in the context of precise Standard-Model (SM)
measurements, but also instrumental for new-physics searches,
especially when looking for small deviations or when providing
cross-section limits on beyond-SM (BSM) signatures.

The last ten years have seen an enormous progress in the development of
hadron-level event generators that include already corrections up to
next-to-next-to-leading order (NNLO) in QCD perturbation theory.
To this end, several methods have been developed to match NNLO QCD
calculations with parton showers (NNLO+PS).
The two main approaches today are the \minnlo{}~\cite{Monni:2019whf,Monni:2020nks} and {\sc Geneva}~\cite{Alioli:2013hqa,Alioli:2021qbf,Gavardi:2023aco} ones. 
Among the existing NNLO+PS methods the \minnlo{} one is a notable
exception, as it
was not only applied to several complex
colour-singlet
processes~\cite{Lombardi:2020wju,Lombardi:2021rvg,Lombardi:2021wug,Buonocore:2021fnj,Zanoli:2021iyp,Gavardi:2022ixt,Haisch:2022nwz,Lindert:2022qdd},
but it was also extended to the case of heavy-quark pair production
\cite{Mazzitelli:2020jio,Mazzitelli:2021mmm,Mazzitelli:2023znt},
i.e.\ processes with colour charges both in the initial and in the
final state.
Nevertheless, there are several other classes of processes for which NNLO+PS predictions are becoming absolutely crucial 
in view of the vast progression of the LHC data taking by the experiments. One important class of processes is the one 
that includes light jets in the final state, such as colour-singlet plus jet production. Although NNLO predictions have been known for
this type of processes for several years, in particular for Higgs+jet \cite{Boughezal:2015dra,Boughezal:2015aha,Caola:2015wna,Chen:2016zka}, $V$+jet \cite{Boughezal:2015dva,Gehrmann-DeRidder:2017mvr,Ridder:2015dxa,Boughezal:2015ded,Campbell:2017dqk}, $VH$+jet \cite{Gauld:2021ule} ($V$ being a vector boson), and 
even $\gamma\gamma$+jet \cite{Chawdhry:2021hkp}, the consistent combination with parton showers is still an open problem.

Current NNLO+PS approaches, such as \minnlo{} \cite{Monni:2019whf,Monni:2020nks} or {\sc Geneva} \cite{Alioli:2013hqa,Alioli:2021qbf,Gavardi:2023aco}, rely on a suitable jet-resolution variable to separate
events at the respective Born level from those with one extra radiation, similar to those NNLO approaches that perform a slicing 
in such jet-resolution variable to reach NNLO accuracy, like $q_T$ subtraction \cite{Catani:2007vq} or $N$-jettiness subtraction \cite{Boughezal:2015eha,Gaunt:2015pea}, or the more recent subtraction based on $p_T$-veto \cite{Abreu:2022zgo}. In order to formulate an NNLO(+PS) method, the resummation ingredients of the jet-resolution variable need
to be known to a sufficiently large order in the logarithmic expansion.
In principle, there is a suitable jet-resolution variable for colour-singlet plus jet processes, namely $1$-jettiness ($\Tau_1$), for which 
a factorization theorem has been derived using
soft-collinear effective theory (SCET)~\cite{Bauer:2000ew, Bauer:2000yr, Bauer:2001ct, Bauer:2001yt, Bauer:2002nz} and whose resummation is know at N$^3$LL~\cite{Alioli:2023rxx}, using the resummation ingredients obtained in \citeres{Campbell:2017hsw,Gaunt:2015pea,Gaunt:2014cfa,Gaunt:2014xga,Bell:2023yso}. In fact, $\Tau_1$ has already been used
to obtain NNLO predictions through slicing
for this class of processes \cite{Boughezal:2015dra,Boughezal:2015aha,Boughezal:2015dva,Boughezal:2015ded,Campbell:2017dqk}.
Although the use of jettiness-like observables in the context of a fully exclusive 
Monte-Carlo simulation poses some challenges, especially related to preserving the parton shower accuracy, at the moment $\Tau_1$ is the only observable for which all the resummation ingredients for an NNLO+PS 
matching are currently available. Therefore, it provides 
our best chance to obtain NNLO+PS predictions for colour-singlet plus jet production in a relatively short time period.

As a first step, it is instructive to derive an NNLO+PS approach
for inclusive colour-singlet production based on
the $0$-jettiness ($\Tau_0$) variable.
In this paper, we therefore
reformulate the \minnlo{} method in terms of $\Tau_0$ as an alternative to the transverse momentum of the colour-singlet final state ($\pt{}$),
and obtain a new implementation for Higgs-boson and vector-boson production at NNLO+PS accuracy.
Our results can be compared directly to the previous \minnlo{} generators  that were built using $\pt{}$ resummation~\cite{Monni:2019whf,Monni:2020nks}. We will show that, despite the fact that for \minnlo{}-$\Tau_0$ the leading logarithmic accuracy in our showered predictions is not fully preserved, we find remarkable agreement for vector-boson production 
before and after including parton-shower effects. For Higgs-boson production, on the other hand, only NNLO quantities, i.e.\ the Higgs rapidity,
are in good agreement, while we find larger differences in jet-related quantities (formally described at NLO QCD accuracy).
In addition, we also compare our predictions to results obtained using the {\sc Geneva} method (both for $\Tau_0$ and $\pt{}$), finding
very good agreement in all NNLO quantities.
The development of a \minnlo{} framework for $\Tau_1$
closely parallels our derivation for $\Tau_0$, and we provide the
corresponding derivation and formulae in \app{sec:higgsplusjet}. The result for $\Tau_1$ introduces
additional complications due to the extra radiation, which generates
logarithmic terms that are absent in the cases of $\Tau_0$ or $\pt$.

This paper is organized as follows: First, we briefly recall the \minnlo{} formalism based on $\pt{}$ in \sct{sec:minnlo}. 
Then, we derive a new formulation of \minnlo{} using $\Tau_0$ in \sct{sec:H0jtau0}, where we review the factorization/resummation 
formula for $\Tau_0$ (\sct{sec:tau0_review}), recast it into a form suitable for the \minnlo{} approach (\sct{sec:derivation_MiNNLO_Tau0}), obtain the \minnlo{} master formula for $\Tau_0$ (\sct{sec:tau0master}) and discuss the shower accuracy (\sct{sec:truncshower}). In \sct{sec:pheno}, after presenting our setup (\sct{sec:setup}), we compare our new
\minnlo{}-$\Tau_0$ results against \minnlo{}-$\pt{}$ predictions (\sct{sec:tau0vspt}) for vector-boson and Higgs-boson production, 
study the difference between \minnlo{} and {\sc Geneva} predictions using both resummation variables for vector-boson production (\sct{sec:MiNNLOvsGeneva}) and present a comparison to ATLAS and CMS data (\sct{sec:data}).
We summarize our findings in \sct{sec:conclusions}.
In \app{app:constants}
all relevant resummation ingredients for $\Tau_0$ are provided,
in \app{app:spreading} we discuss how higher-order resummation terms are spread over the inclusive phase space, 
%in \app{app:H0jptSCET} the \minnlo{}-$\pt{}$ method is rederived starting from a SCET formulation,
 and 
in \app{sec:higgsplusjet} we present  the derivation of the \minnlo{} approach for $\Tau_1$.

%=======================================
\section{\minnlo{} in a nutshell}
\label{sec:minnlo}
The \minnlo{} method~\cite{Monni:2019whf,Monni:2020nks} formulates a fully differential calculation
in the Born phase space $\PhiB$ at NNLO QCD accuracy of a produced final state 
 \F{} with invariant mass $Q$ and combines it consistently with multi-parton radiation
effects from parton shower.  \minnlo{} has been formulated thus far for colour-singlet
production \cite{Monni:2019whf,Monni:2020nks} and for heavy-quark pair production \cite{Mazzitelli:2020jio,Mazzitelli:2021mmm}, based
on the transverse momentum ($\pt$) of \F{} as a matching variable in both cases. Here, we focus on the
colour-singlet case and, when specific to the original formulation based on $\pt$ resummation, we will refer to it as \minnlo{}-$\pt{}$ from now on.
 
\minnlo{} starts from a differential description of the production of the colour singlet and a jet (\FJ{})
at NLO matched to parton showers, which is obtained via the \POWHEG{} approach \cite{Nason:2004rx,Frixione:2007vw,Alioli:2010xd} as
\begin{align}
\frac{\mathd\sigma}{\mathd\PhiBJ}={\bar B}(\PhiBJ) \times
\bigg\{\Delta_{\rm pwg} (\LambdaPWG) + \int\mathd \Phi_{\tmop{rad}} 
  \Delta_{\rm pwg} (\ptrad)  \frac{R (\PhiBJ{}, \Phi_{\tmop{rad}})}{B
  (\PhiBJ{})}\bigg\}\,,
\label{eq:master}
\end{align}
where \minnlo{} modifies the content of the ${\bar B}(\PhiBJ)$ function,
which generates the first radiation (inclusive over the second one),
to achieve NNLO QCD accuracy. By contrast, the
content of the curly brackets, which describes the exclusive generation of the second
radiation according to the \POWHEG{} method, is not modified by the \minnlo{} procedure. 
Here, $\PhiBJ$ denotes the  \FJ{} phase space, and $B$ and $R$ are the squared tree-level matrix elements for \FJ{}
and \FJJ{} production, respectively.  
$\Delta_{\rm pwg}$ is the usual \POWHEG{} Sudakov form factor~\cite{Nason:2004rx} (with a lower cutoff of $\LambdaPWG=0.89$~GeV) and
the phase space (transverse momentum) of the second radiation is denoted as $ \Phi_{\tmop{rad}} $ ($\ptrad$). 
Extra radiation beyond the second one is added by the consistent matching to the parton shower achieved through \POWHEG{}.

The modification of the ${\bar B}(\PhiBJ)$ function is the central ingredient of
\minnlo{}. For \minnlo{}-$\pt$, its derivation \cite{Monni:2019whf} stems from the
description of the NNLO cross section differential in the $\pt{}$ of the colour singlet 
and in the Born phase space $\PhiB$, which can be expressed as
\begin{align}
\label{eq:minnloptstart}
  \frac{\mathd\sigma}{\mathd\PhiB\mathd \pt} &= \frac{\mathd}{\mathd \pt}
     \bigg\{ \exp[-\tilde{S}(\pt)] \cL(\pt)\Bigg\} +
                                               R_f(\pt) =
  \exp[-\tilde{S}(\pt)]\left\{
                                  D(\pt)+\frac{R_f(\pt)}{\exp[-\tilde{S}(\pt)]}\right\}\,,
\end{align}
where $R_f$ contains the non-singular terms in the $\pt\rightarrow 0$ limit, and 
\begin{equation}
\label{eq:Dterms_pT}
  D(\pt)  \equiv -\frac{\mathd \tilde{S}(\pt)}{\mathd \pt} \cL(\pt)+\frac{\mathd \cL(\pt)}{\mathd \pt}\,.
\end{equation}
$\tilde{S}(\pt)$ represents the Sudakov form factor in $\pt$, while $\cL(\pt)$
contains the parton luminosities, the squared hard matrix elements
for the underlying \F{} production process up to two loops as well as
the collinear coefficient functions at NNLO (see Ref.~\cite{Monni:2019whf} for further
details). 
A crucial feature of the \minnlo{}-$\pt$ procedure is that the renormalization
and factorization scales are set to $\muR\sim\muF\sim \pt$.

Let us introduce the NLO differential cross section for \FJ{} production as 
\begin{equation}
\label{eq:NLO}
\frac{\mathd\sigma^{\rm (NLO)}_{\scriptscriptstyle\rm FJ}}{\mathd\PhiB\mathd
      \pt} = \abarmu{\pt}\left[\frac{\mathd\sigma_{\scriptscriptstyle\rm FJ}}{\mathd\PhiB\mathd
      \pt}\right]^{(1)} + \left(\abarmu{\pt}\right)^2\left[\frac{\mathd\sigma_{\scriptscriptstyle\rm FJ}}{\mathd\PhiB\mathd
      \pt}\right]^{(2)}\,,
\end{equation}
with $[X]^{(i)}$ being the coefficient of the
$i$-th term in the perturbative expansion of the quantity $X$,
so that we can write the non-singular terms as
\begin{align}
R_f = \frac{\mathd\sigma^{\rm (NLO)}_{\scriptscriptstyle\rm FJ}}{\mathd\PhiB\mathd
      \pt} - \abarmu{\pt} \left[\exp[-\tilde{S}(\pt)] D(\pt)\right]^{(1)} - \left(\abarmu{\pt}\right)^2 \left[\exp[-\tilde{S}(\pt)] D(\pt)\right]^{(2)}\,.
\end{align}
Now, we can rewrite \eqn{eq:minnloptstart} as
\begin{align}
\label{eq:minnlopT}
  \frac{\mathd\sigma}{\mathd\PhiB\mathd \pt}  &=
  \exp[-\tilde{S}(\pt)]\bigg\{ \abarmu{\pt}\left[\frac{\mathd\sigma_{\scriptscriptstyle\rm FJ}}{\mathd\PhiB\mathd
      \pt}\right]^{(1)} \left(1+\abarmu{\pt} [\tilde{S}(\pt)]^{(1)}\right)
  \notag \\
& + \left(\abarmu{\pt}\right)^2\left[\frac{\mathd\sigma_{\scriptscriptstyle\rm FJ}}{\mathd\PhiB\mathd
      \pt}\right]^{(2)} \notag + \left[D(\pt) -\abarmu{\pt} [D(\pt)]^{(1)}
  -\left(\abarmu{\pt}\right)^2 [D(\pt)]^{(2)}  \right] \\ & +  {\rm
  regular~terms~of~{\cal O}(\as^3)}\bigg\}\,.
\end{align}
This formula includes all relevant terms needed to reach NNLO QCD accuracy. In particular,
upon integration over $\pt$ from scales of the order of the Landau pole
$\Lambda$ to the kinematic upper bound, it reproduces the fully differential NNLO cross section
up to terms beyond accuracy.

Each term of \eqn{eq:minnlopT} contributes to the integrated cross section with scales
$\muR\sim\muF\sim Q$ according to the power counting formula
\begin{equation}
    \int_{\Lambda}^{Q} \mathd \pt \frac{1}{\pt} \as^m(\pt) \ln^n\frac{Q}{\pt}
\exp(-\tilde{S}(\pt))    \approx {\cal O}\left(\as^{m-\frac{n+1}{2}}(Q)\right)\,.
\end{equation}
Here, it is crucial to understand that for \minnlo{}-$\pt$ at most an extra single logarithm
appears in the formula due to the choice $\muR\sim\muF\sim \pt$, therefore $n\le 1$.
As a consequence, the truncation of \eqref{eq:minnlopT} beyond second order, i.e.\ all terms including those up to $\as^2(\pt)$, is NLO accurate
in the $\PhiB$ phase space, which corresponds to \minlo{} \cite{Hamilton:2012rf} accuracy, while the terms at third order in $\as(\pt)$ and beyond 
in the square brackets are the crucial ingredients to reach NNLO accuracy in $\PhiB$.
Moreover, one can expand the square bracket in the second line of \eqn{eq:minnlopT} and neglect terms that produce N$^3$LO corrections or beyond upon integration over
$\pt$, to any inclusive observable in $\PhiB$. 
One can therefore, in principle, truncate the second line of \eqn{eq:minnlopT} to third order in $\as(\pt)$
\begin{align}
\begin{split}
\label{eq:truncated_D}
D^{(\ge 3)}(\pt) &= 
D(\pt) -\abarmu{\pt} [D(\pt)]^{(1)} -\left(\abarmu{\pt}\right)^2
[D(\pt)]^{(2)} \\
&=\left(\abarmu{\pt}\right)^3 [D(\pt)]^{(3)} +{\cal O}(\as^4(\pt))\,.
\end{split}
\end{align}
Nevertheless, in order to preserve the full total derivative of our 
starting equation \eqn{eq:minnloptstart}, and to keep terms beyond accuracy generated by the total derivative,
it is preferable not to perform such a truncation, although formally valid, as pointed out in \citere{Monni:2020nks}.

The \minnlo{}-$\pt$ procedure can be applied directly at the fully differential level
in the $\PhiBJ$ phase space to the ${\bar B}(\PhiBJ)$ function in \eqn{eq:master}~\cite{Monni:2019whf}:
\begin{align}
\label{eq:Bbar}
{\bar B}(\PhiBJ)&\equiv \exp[-\tilde{S}(\pt)]\bigg\{ \abarmu{\pt}\left[\frac{\mathd\sigma_{\scriptscriptstyle\rm FJ}}{\mathd\PhiBJ}\right]^{(1)} \left(1+\abarmu{\pt} [\tilde{S}(\pt)]^{(1)}\right)\notag
  \\
&+ \left(\abarmu{\pt}\right)^2\left[\frac{\mathd\sigma_{\scriptscriptstyle\rm FJ}}{\mathd\PhiBJ}\right]^{(2)} + D^{(\ge 3)}(\pt)\,  F^{\tmop{corr}}(\PhiBJ)\bigg\}\,,
\end{align}
where the factor $F^{\tmop{corr}}(\PhiBJ)$ encodes a suitable function needed to spread the correction $D^{(\ge 3)}(\pt)$,
which intrinsically depends only on $\pt$ and $\PhiB$, on the full $\PhiBJ$ phase space, as discussed 
in detail in Section 3 of \citere{Monni:2019whf}.

%----------------------------------------------------------------------------
\section{Formulation of the \minnlo{} method using $0$-jettiness}
\label{sec:H0jtau0}

In this section, we derive the \minnlo{} master formula for colour-singlet production 
using $\Tau_0$ factorization and resummation as
formulated in SCET
From now on, we will refer to this new formulation of the \minnlo{} method as \minnlo{}-$\Tau_0$.
Our derivation follows the strategy of the \minnlo{} method based on $\pt{}$.
%For convenience of the reader, we also briefly present the corresponding derivation
%based on $\pt{}$ factorization using SCET in \app{app:H0jptSCET}.
In order to perform the \minnlo{}-$\Tau_0$ derivation we start by reviewing 
$\Tau_0$ factorization and resummation as formulated in SCET in \sct{sec:tau0_review}.
Then, in \sct{sec:derivation_MiNNLO_Tau0}, we bring it into a form that makes it 
suitable to make contact with the \minnlo{} approach and
we derive the \minnlo{}-$\Tau_0$ master formula in \sct{sec:tau0master}.
Finally, we discuss the accuracy of the parton shower in \sct{sec:truncshower}.

% For clarity, we will use the following notations:
% \begin{align}
%  \frac{\df\sigma^{\rm sing}}{\df\PhiB \df \Tau_0} =
%  \int\frac{\df y}{2\pi} \, e^{\img y \Tau_0} \cL(y_0/y) e^{-\cS(y_0/y)}
% \end{align}
% for the $\Tau_0$ spectrum canonically resummed in Fourier space, and
% \begin{align}
%  \frac{\df\sigma^{\rm sing}(\TauCut)}{\df\PhiB} = \tilde\cL(\TauCut) e^{-\tilde \cS(\TauCut)} + \cO(\as^3)
% \end{align}
% for the cumulant with momentum-space scale setting following the \minnlo{} method, which is only accurate at NNLO in $\as$.
% (Note that we use $\cS$ for the Sudakov to distinguish it from the soft function.)

%----------------------------------------------------------------------------
\subsection{Review of $\Tau_0$ factorization and evolution}
\label{sec:tau0_review}
% %%%%%%%%%%%%%%%%%%%%%%%%%%%%%%%%%%%%%%%%%%%%%%%%%%%%%%%%%%%%%%%%%%%%%%%%%%%%%

We begin by reviewing the SCET resummation formalism for $0$-jettiness ($\Tau_0$), also known as beam thrust.
The infrared-safe $0$-jet resolution variable $\Tau_0$ was originally introduced in \citere{Stewart:2009yx},
and extended to $N$-jet processes in \citere{Stewart:2010tn}. Following the notation of \citere{Jouttenus:2011wh},
it is defined as
\begin{align} \label{eq:def_Tau0}
 \Tau_0 = \sum_k \min \biggl\{ \frac{2 q_a \cdot p_k}{Q_a} \,, \frac{2 q_b \cdot p_k}{Q_b} \biggr\}
\,.\end{align}
Here, the sum runs over all hadronic final states (excluding decay products
of the identified colour-singlet final state), $p_k$ are their momenta,
and $q_a$ and $q_b$ are the momenta of the colliding partons.
Aligning the incoming hadrons along the directions
\begin{align}
 n^\mu = (1, 0, 0, 1)
\,,\qquad
 \bn^\mu = (1, 0, 0, -1)
\,,\end{align}
the incoming parton momenta can be written as
\begin{align}
 q_a^\mu = Q \, e^Y \, \frac{n^\mu}{2}
\,,\qquad
 q_b^\mu = Q \, e^{-Y} \, \frac{\bn^\mu}{2}
\,,\end{align}
where $Q$ and $Y$ are the invariant mass and rapidity of the colour-singlet final state, respectively.
Finally, $Q_{a,b}$ in \eq{def_Tau0} are normalization factors giving rise to different definitions of $\Tau_0$.
The most common choices are~\cite{Stewart:2009yx, Stewart:2010tn}
\begin{alignat}{3}
 &\text{leptonic}~\Tau_0: \qquad && Q_a = Q_b = Q \,, \qquad &&
   \Tau_0^{\rm lep} = \sum_k \min \bigl\{ e^{+Y} n \cdot p_k \,, e^{-Y} \bn \cdot p_k \bigr\}
\,,\nn\\
  &\text{hadronic}~\Tau_0: \qquad && Q_{a,b} = Q e^{\pm Y} \,, \qquad &&
    \Tau_0^{\rm cm} = \sum_k \min \bigl\{ n \cdot p_k \,, \bn \cdot p_k \bigr\}
 %&\text{hadronic}~\Tau_0: \qquad && Q_a = Q_b = Q e^{\pm Y} \,, \qquad &&
 %  \Tau_0^{\rm had} = \sum_k \min \bigl\{ n \cdot p_k \,, \bn \cdot p_k \bigr\}
\,.\end{alignat}
It has been shown that the power corrections to the hadronic definition are exponentially enhanced in $Y$~\cite{Moult:2016fqy, Moult:2017jsg, Ebert:2018lzn}.
This is compensated for in the leptonic definition by the explicit $e^{\pm Y}$ factors,
which is thus the preferred choice and which we will use throughout this paper.

In the kinematic limit $\Tau_0 \to 0$, all hadronic momenta $p_k$ must be either soft
or collinear to the incoming partons to yield a negligible contribution to the sum in \eq{def_Tau0}.
Based on this observation, a factorization formula%
\footnote{Starting at $\cO(\as^4)$, the factorization is violated by Glauber
contributions~\cite{Gaunt:2014ska}.}
was derived in \citeres{Stewart:2009yx, Stewart:2010tn} using
SCET.
It can be written as
\begin{align} \label{eq:Tau0_fact}
 \frac{\df\sigma}{\df\PhiB \df\Tau_0} &
 = \frac{\df\sigma^{\rm sing}}{\df\PhiB \df\Tau_0} \times \left[ 1 + \cO\bigl(\sqrt{\Tau_0/Q}\bigr)\right]
\,,\\\nn
 \frac{\df\sigma^{\rm sing}}{\df\PhiB \df\Tau_0} &
 = \sum_{a,b} \frac{\df |\cM_{ab}|^2}{\df \PhiB} H_{ab}(Q, \mu) \int\!\df t_a \df t_b \,
   B_a(t_a, x_a, \mu) B_b(t_b, x_b, \mu) S\Bigl( \Tau_0 - \frac{t_a}{Q_a} - \frac{t_b}{Q_b}, \mu \Bigr)
\,,\end{align}
where we are differential both in $\Tau_0$ and in the Born phase space $\PhiB$.
As indicated, the factorization holds up to power corrections in $\sqrt{\Tau_0/Q}$, see \citere{Ebert:2019zkb} for instance.
The sum runs over all flavour combinations $a$ and $b$ contributing to the Born process,
where $\cM_{ab}$ is the corresponding matrix element and the hard function $H_{ab}$ encodes virtual
corrections to it. The beam functions $B_{a,b}$ encode the effect of radiation
close the incoming protons, and they are convolved (integral over $t_a$ and $t_b$) against the soft function $S$ encoding soft radiation.
The soft function differs between quark and gluon-induced processes, which is kept implicit in \eqn{eq:Tau0_fact}, 
but it is independent of the quark flavour in the massless case. 
Finally, $x_{a,b} = Q e^{\pm Y}/\Ecm$ are the Bjorken variables in the Born kinematics,
and the renormalization and factorization scales are denoted by $\mu$.
Note that all perturbative ingredients depend on the process under
consideration (quark versus gluon initiated), but the form of the
equations is the same, therefore we leave this dependence
implicit.\footnote{In the case of $\pt{}$ factorization there are additional collinear correlation functions~\cite{Catani:2010pd} for gluon-initiated processes starting at NNLO, which are
  absent in the case of $\Tau_0$ due to the scalar nature of this
  observable.}

While it is standard in the literature to discuss $\Tau_0$ in momentum space,
here we perform a Fourier transform with respect to $y = y - \img0$
to turn the convolution in \eq{Tau0_fact} into a simple product,
similar to the usual treatment in $\pt{}$ factorization.
This yields
\begin{align} \label{eq:Tau0_fact_FT}
 \frac{\df\sigma^\sing}{\df\PhiB \, \df\Tau_0} &
 = \sum_{a,b} \frac{\df |\cM_{ab}|^2}{\df \PhiB} H_{ab}(Q, \mu)
   \int\frac{\df y}{2\pi} \, e^{\img y \Tau_0}
   B_a\Bigl(\frac{y}{Q_a},x_a,\mu\Bigr) B_b\Bigl(\frac{y}{Q_b},x_b,\mu\Bigr) S(y, \mu)
\,,\end{align}
where the Fourier transformed beam and soft functions are defined as
\begin{alignat}{3} \label{eq:FT}
 B_j(y, x, \mu) &= \int\df t \, e^{-\img t y} B_j(t, x, \mu)
\,,\qquad
 S(y, \mu) &= \int\df \Tau \, e^{-\img \Tau y} S(\Tau, \mu)
\,.\end{alignat}
Note that the arguments of $B_j$ and $S$ have different mass dimensions due to geometric measures,
which in \eq{Tau0_fact_FT} we have put back into the arguments of the beam functions.

In Fourier space, the hard, beam and soft functions obey the following renormalization group equations (RGEs):%
\footnote{For processes such as Higgs production, $\cM_{ab}$ also carries a scale dependence,
which would be compensated by changing $\gamma_H$ accordingly.
Alternatively, one can always simultaneously evolve $|\cM_{ab}|^2$ and $H_{ab}$,
which leaves $\gamma_H$ unchanged.}
\begin{alignat}{3} \label{eq:Tau0_RGEs}
 \frac{\df}{\df\ln\mu} \ln H_{ab}(Q, \mu) &
 = \gamma_H(Q^2, \mu)
 &&= \phantom{-}2 \GammaC[\as(\mu)] \ln\frac{Q^2}{\mu^2} &&+ \gamma_H[\as(\mu)]
\,,\nn\\
 \frac{\df}{\df\ln\mu} \ln B_i\Bigl(\frac{y}{Q_i},x,\mu\Bigr) &
 = \gamma_B\Bigl(\frac{y}{Q_i},\mu\Bigr)
 &&= \phantom{-}2 \GammaC[\as(\mu)] \ln\frac{y \mu^2}{y_0 Q_i} &&+ \gamma_B[\as(\mu)]
\,,\nn\\
 \frac{\df}{\df\ln\mu} \ln S(y,\mu) &
 = \gamma_S(y, \mu)
 &&= -4 \GammaC[\as(\mu)] \ln\frac{y \mu}{y_0} &&+ \gamma_S[\as(\mu)]
\,.\end{alignat}
Here, $\GammaC$ is the cusp anomalous dimension, and $\gamma_{H,B,S}$
are the hard, beam and soft non-cusp anomalous dimensions, respectively.
For brevity, we also defined the constant
\begin{align} \label{eq:def_y0}
 y_0 = -\img e^{-\gamma_E}
\end{align}
appearing in the logarithms, where $\gamma_E$ is the Euler-Mascheroni constant. Note that due to the Fourier transform,
the beam anomalous dimension explicitly depends on the measure $Q_i$,
which in momentum space only appears in the convolution in \eq{Tau0_fact}.
Finally, the overall $\mu$ independence implies that
\begin{align} \label{eq:RGE_consistency}
 \gamma_H(\as) + 2 \gamma_B(\as) + \gamma_S(\as) & = 0
\,,\end{align}
as well as $Q_a Q_b = Q^2$, which is fulfilled for both hadronic and leptonic $\Tau_0$.

It is clear from \eqn{eq:Tau0_RGEs} that the hard, beam and soft functions contain large logarithms
that can be minimized using the scale choices
\begin{align} \label{eq:canonical_scales_Tau0}
 \mu_H = Q
\,, \qquad
 \mu_B = \sqrt{\frac{Q y_0}{y}} = \sqrt{\mu_H \mu_S}
\,, \qquad
 \mu_S = \frac{y_0}{y}
\,,\end{align}
for $H$, $B$ and $S$, respectively. Note that for generic measures $Q_a \ne Q_b$, this would leave potentially
large logarithms $\ln(Q/Q_i)$ in the individual beam functions $B_i$,
which however is absent for our preferred choice $\Tau_0^{\rm lep}$.
By solving \eq{Tau0_RGEs}, we can evolve all functions appearing in \eq{Tau0_fact_FT}
from their natural scales in \eq{canonical_scales_Tau0} to a common scale $\mu$,
which resums large logarithms $\ln(Q/\Tau_0)$ appearing in the cross section.
The resummed cross section is then given by
\begin{align} \label{eq:Tau0_resummed_1}
 \frac{\df\sigma^\sing}{\df\PhiB \, \df\Tau_0} &
 = \int\frac{\df y}{2\pi} \, e^{\img y \Tau_0} \,
   \sum_{a,b}  \frac{\df |\cM_{ab}|^2}{\df \PhiB} H_{ab}(Q, \mu_H) B_a\Bigl(\frac{y}{Q_a}, x_a, \mu_B \Bigr) B_b\Bigl(\frac{y}{Q_b}, x_b, \mu_B \Bigr) S(y, \mu_S)
   \nn\\&\quad\times
   \exp\left\{ - \int_{\mu_B}^{\mu_H} \frac{\df\mu'}{\mu'} \left[ 2 \GammaC[\as(\mu')] \ln\frac{Q^2}{\mu'^2} + \gamma_H[\as(\mu')] \right] \right\}
   \nn\\&\quad\times
   \exp\left\{ - \int_{\mu_B}^{\mu_S} \frac{\df\mu'}{\mu'} \left[ -4 \GammaC[\as(\mu')] \ln\frac{y \mu'}{y_0} + \gamma_S[\as(\mu')] \right] \right\}
\,,\end{align}
where the two exponentials evolve the hard and soft functions from their natural scales $\mu_H$ and $\mu_S$ to the beam scale $\mu_B$, respectively.

To bring this into a form suitable for the \minnlo{} method,
we first notice that
for the hard and soft functions, using their natural scales, the coefficients in the perturbative expansion are constant\footnote{For $2\to2$ or more complicated processes, the hard function constants
actually have a nontrivial dependence on $\PhiB$, which we keep implicit.}, i.e.
%the hard and soft functions reduce to pure constants at their natural scales,\commentgz{rephrase? e.g. y not constant... mention dependence only through running..?}%
\begin{alignat}{2} \label{eq:H_S_canonical}
 H_{ab}(Q)
 &\equiv H_{ab}(Q, \mu_H = Q) %\equiv H[\as(Q)] \equiv H(Q)
 &&= \sum_{n=0}^\infty \left[\frac{\as(Q)}{2\pi}\right]^n H_{ab}^{(n)}
\,,\nn\\
 S(y_0/y)
 &\equiv S(y, \mu_S = y_0/y) %\equiv S[\as(y_0/y)] \equiv S(y_0/y)
 &&= \sum_{n=0}^\infty \left[\frac{\as(y_0/y)}{2\pi}\right]^n S^{(n)}
\,.\end{alignat}
We can then express the first line
in \eq{Tau0_resummed_1} at a common scale, as used in the \minnlo{} method. 
Since the beam scale $\mu_B$ enters the nonperturbative parton distribution functions (PDFs),
it is natural to choose $\mu_B$ as this common scale. Thus, with some abuse of notation 
($H_{ab}(Q) \equiv H_{ab}[\as(Q)]$ and $S(y_0/y) \equiv S[\as(y_0/y)]$), we rewrite \eq{H_S_canonical} as
\begin{align} \label{eq:shifted_H_S}
 H_{ab}(Q) &
 = H_{ab}(\mu_B) \exp\left[ \int_{\mu_B}^{Q} \frac{\df\mu'}{\mu'} \gamma_{\bar H}[\as(\mu')] \right]
\,,\nn\\
 S(y_0/y) &
 = S(\mu_B) \exp\left[ \int_{\mu_B}^{y_0/y} \frac{\df\mu'}{\mu'} \gamma_{\bar S}[\as(\mu')] \right]
\,,\end{align}
where the induced anomalous dimensions are given by
\begin{align}
 \gamma_{\bar F}(\as) &
 = 2 \beta(\as) \frac{\df\ln F(\as)}{\df\as}
 = -4 \pi \beta_0 \frac{F^{(1)}}{F^{(0)}} \Bigl(\frac{\as}{2\pi}\Bigr)^2 + \cO(\as^3)\,,
\end{align}
for $F=H_{ab}$ or $F=S$. $\beta(\as)$ represents the QCD $\beta$ function
as defined in \eqref{eq:conventions_2}, and $\beta_0$ it's first order coefficient given in \eqref{eq:beta_coeffs}.
The beam functions can be related to the PDFs as
\begin{align} \label{eq:beam_matching}
 B_i\Bigl(\frac{y}{Q_i}, x, \mu\Bigr) &
 = \sum_j \int_x^1 \frac{\df x'}{x'} \bar C_{ij}\Bigl(\frac{y}{Q_i}, x', \mu\Bigr) f_j\Bigl(\frac{x}{x'}, \mu\Bigr)
 \equiv (\bar C \otimes f)_i\Bigl(\frac{y}{Q_i}, x, \mu\Bigr)
\,,\end{align}
where the $\bar C_{ij}$ are perturbatively calculable coefficients.
At their canonical scale, they read %they can be written as
\begin{alignat}{2} \label{eq:beam_constants}
 \bar C_{ij}\Bigl(\frac{y}{Q_i}, x\Bigr)
 &\equiv
 \bar C_{ij}\Bigl(\frac{y}{Q_i}, x, \mu_B \Bigr)
 &&= \sum_{n=0}^\infty \left[\frac{\as(\mu_B)}{2\pi}\right]^n \bar C_{ij}^{(n)}(x)
\,,\end{alignat}
with all dependence on $y$ arising through $\as(\mu_B)$.
Similar to the treatment commonly used in $\pt{}$ resummation, where the beam and soft functions are combined into a transverse-momentum dependent PDF (TMDPDF),
we absorb the $\Tau_0$ soft function with the beam function, and define
\begin{align} \label{eq:beam_matching_2}
 C_{ij}\Bigl(\frac{y}{Q_i}, x\Bigr) \equiv \bar C_{ij}\Bigl(\frac{y}{Q_i}, x, \mu_B \Bigr) \sqrt{S(\mu_B)}
\,.\end{align}
Using \neqn{eq:shifted_H_S}--\eqref{eq:beam_matching_2} to rewrite \eq{Tau0_resummed_1}, we arrive at our final expression
\begin{align} \label{eq:Tau0_resummed}
 \frac{\df\sigma^\sing}{\df\PhiB \, \df\Tau_0} &
 = \int\frac{\df y}{2\pi} \, e^{\img y \Tau_0} \, \cL(y_0/y) \, e^{-\cS(y_0/y)}
\,,\end{align}
where the luminosity and Sudakov factor are defined as
\begin{align} 
 \cL(y_0/y) &
 = \sum_{a,b}
   \frac{\df |\cM_{ab}|^2}{\df \PhiB} H_{ab}(\mu_B) \,
   (C \otimes f)_a(x_a, \mu_B) \,
   (C \otimes f)_b(x_b, \mu_B)
%%%
\,,\nn\\
%%%
 \cS(y_0/y) &
 = 2 \int_{\mu_B}^{\mu_H} \frac{\df\mu'}{\mu'} \left[ A[\as(\mu')] \ln\frac{Q^2}{\mu'^2} + B_H[\as(\mu')] \right] \label{eq:Tau0_L_S}
 \\&
 + 2 \int_{\mu_B}^{\mu_S} \, \frac{\df\mu'}{\mu'} \left[ A[\as(\mu')] \ln\frac{(y_0/y)^2}{\mu'^2} + B_S[\as(\mu')] \right]\nn
 \,.\end{align}
In what follows, we will keep the dependence of the PDFs on the momentum fractions
$x_a$ and $x_b$ and on $\mu_B$ implicit.
Following the standard naming conventions in \minnlo, the anomalous dimensions are labeled as $A$ and $B$.
They are related to the anomalous dimensions in \eq{Tau0_RGEs} by
\begin{align} \label{eq:def_B_Tau0}
 A(\as) = \GammaC(\as)
\,,\quad
 B_{F}(\as) = \frac12 \gamma_F(\as) - \beta(\as) \frac{\df\ln F(\as)}{\df\as}
\,,\end{align}
where $F=H$ for the hard function and $F=S$ for the soft function.
Note that the structure of $\cS$ contains two distinct evolution kernels,
and as such its structure differs from the Sudakov for $\pt{}$ resummation,
which can be written as a single kernel. %, cf.~\eq{qT_S_2}.

We expand the coefficients in \eq{def_B_Tau0} as
\begin{align}
 A(\as) = \sum_{n=1}^\infty A^{(n)} \left(\frac{\as}{2\pi}\right)^n
\,,\qquad
 B_F(\as) = \sum_{n=1}^\infty B_F^{(n)} \left(\frac{\as}{2\pi}\right)^n
\,.\end{align}
Explicit results for all required coefficients are collected in \app{app:constants}.

%===============================================================================
\subsection{Making contact with the \minnlo{} method}
\label{sec:derivation_MiNNLO_Tau0}
%===============================================================================

We have already brought the $\Tau_0$ resummation formula into a form that is suitable 
to make contact with the \minnlo{} method, see \eqs{Tau0_resummed}{Tau0_L_S}.
We begin by taking the cumulant of \eq{Tau0_resummed},
\begin{align} \label{eq:TauCut_resummed_1}
 \frac{\df \sigma^{\rm sing}(\TauCut)}{\df\PhiB} &
 = \int_0^{\TauCut} \df\Tau_0 \int\frac{\df y}{2\pi} \, e^{\img y \Tau_0} \, \cL(y_0/y) \, e^{-\cS(y_0/y)}
\,,\end{align}
which we expand around $y_0 / y \sim \TauCut$.
More precisely, we define our expansion through
\begin{align} \label{eq:def_Ly}
 L_y = \ln\frac{\TauCut y}{y_0} \ll 1
\,.\end{align}
Expanding the luminosity and Sudakov yields
\begin{align} \label{eq:expansions_cL_S_Tau0}
 \cL(y_0/y) &
 = \cL(\TauCut) + L_y \, \cL'(\TauCut) + \frac12 L_y^2 \, \cL''(\TauCut) + \cdots
\,,\nn\\
 e^{-\cS(y_0/y)} &
 = e^{-\cS(\TauCut) - L_y \, \cS'(\TauCut) } \left[ 1 - \frac12 L_y^2 \, \cS''(\TauCut) - \frac16 L_y^3  \, \cS'''(\TauCut) + \cdots \right]
\,,\end{align}
where the derivatives of an arbitrary function $f$ are defined as
\begin{align} \label{eq:def_fn}
 f^{(n)} &
 = \frac{\df^n f}{\df^n \ln(1/\TauCut)}
 = (-1)^n \frac{\df^n f}{\df^n \ln\TauCut},\quad \textrm{with}\quad f'\equiv f^{(1)}, f''\equiv f^{(2)}, f'''\equiv f^{(3)}, \ldots  
\,.\end{align}
By evaluating the integral of the cumulant using
\begin{align} \label{eq:tau0_cumulant}
 \int_0^{\TauCut} \df \Tau_0 \int \frac{\df y}{2\pi} e^{i \Tau_0 y} L_y^n  e^{-\cS' L_y}
 = (-1)^n \partial^n_{\cS'}  \frac{e^{-\gamma_E \cS'}}{\Gamma(1+\cS')}\,,
\end{align}
and by truncating \eq{TauCut_resummed_1} at NNLO accuracy, we obtain
\begin{align} \label{eq:TauCut_resummed_2}
 \frac{\df \sigma^{\rm sing}(\TauCut)}{\df\PhiB} &
 = e^{-\cS(\TauCut)}  \Bigl[
   \cL(\TauCut) \Bigl(1 - \frac12 \partial_{\cS'}^2 \cS'' + \frac16 \partial_{\cS'}^3 \cS''' + \frac18 \partial_{\cS'}^4 (\cS'')^2 \Bigr)
   \nn\\& \quad
    + \cL'(\TauCut) \Bigl( -\partial_{\cS'} + \frac12 \partial_{\cS'}^3 \cS'' \Bigr)
    + \frac12 \partial_{\cS'}^2 \cL''(\TauCut)
    + \cO(\as^3)
   \Bigr]  \frac{e^{-\gamma_E \cS'}}{\Gamma(1+\cS')}
\,.\end{align}
Here we used that the power counting in terms of the strong coupling constant is given by: 
%the individual derivatives scale as
\begin{align} \label{eq:scalings_derivatives_Tau0}
 \cS' = \cO(\as) \,,\quad \cS'' = \cO(\as) \,,\quad \cS''' = \cO(\as^2) \,,\quad \cL' = \cO(\as) \,,\quad \cL'' = \cO(\as^2)
 \,.\end{align}
%\commentgz{make clear that we are not including logs here, only power of alphas}
Expanding also the Gamma factor in \eq{TauCut_resummed_2} and 
taking the partial derivatives with respect to $\cS'$, we arrive at
\begin{align} \label{eq:TauCut_resummed_3}
 \frac{\df \sigma^{\rm sing}(\TauCut)}{\df\PhiB} &
 = e^{-\cS(\TauCut)}  \Bigl[
   \cL(\TauCut) \Bigl( 1 - \frac{\zeta_2}{2} [(\cS')^2 - \cS''] - \zeta_3 \cS' \cS'' + \frac{3 \zeta_4}{16} (\cS'')^2 + \frac{\zeta_3}{3} \cS''' \Bigr)
   \nn\\&\hspace{2.cm}
   + \cL'(\TauCut) \bigl( \zeta_2 \cS' + \zeta_3 \cS'' \bigr)
   - \frac{\zeta_2}{2} \cL''(\TauCut)
   + \cO(\as^3) \Bigr]
\,.\end{align}
The terms in square brackets are to be understood as an expansion in $\as(\mu_B)$
with $\mu_B = \sqrt{Q \TauCut}$, since $\mu_B$ defined in \eq{canonical_scales_Tau0} has also been expanded around $\TauCut.$

Before proceeding, we note that \eq{TauCut_resummed_3} has a much richer structure
than its counterpart in \minnlo{}-$\pt{}$/ %, see \eq{qTcut_resummed_3}.
%The reason is that \eq{qT_cumulant}, the analog of \eq{tau0_cumulant} in $\pt{}$,
%has a much simpler expansion itself.
The reason is that the analog of \eq{tau0_cumulant} in $\pt{}$
has a much simpler expansion itself.
Most notably, for \minnlo{}-$\pt{}$ one does not
encounter the $\cL''$ term, which will have a significant impact on our final formula,
and there are fewer $\cS'$ terms in the coefficients of $\cL$ and $\cL'$.

Our goal is to write \eq{TauCut_resummed_3} in the form
\begin{align} \label{eq:TauCut_resummed_4}
 \frac{\df \sigma^{\rm sing}(\TauCut)}{\df\PhiB} &
 = \tilde \cL(\TauCut) e^{- \cS(\TauCut)}
\,,\end{align}
which has the same structure as the singular part of the starting formula in \eq{minnloptstart} of the \minnlo{}-$\pt$ method. Thus, bringing the $\Tau_0$ resummation 
formula into that form allows us to follow the same subsequent steps when 
deriving the \minnlo{}-$\Tau_0$ formalism.
Similar to \minnlo{}-$\pt{}$ this can be achieved by absorbing
the additional terms in square brackets in \eq{TauCut_resummed_3} into 
a redefinition of the Sudakov factor and the luminosity. Evaluating the derivatives of $\cS$
in \eq{TauCut_resummed_3}, we obtain the expression
\begin{align} \label{eq:TauCut_resummed_5}
 \frac{\df \sigma^{\rm sing}(\TauCut)}{\df\PhiB}
 = e^{-\cS(\TauCut)}  \Bigl\{ &
     \cL(\TauCut) \Bigl[ 1 + \frac{\as}{2\pi} c_{1,0} + \Bigl(\frac{\as}{2\pi}\Bigr)^2 \bigl( c_{2,2} L_\Tau^2 + c_{2,1} L_\Tau + c_{2,0} \bigr) \Bigr]
   \nn\\&
   + \cL'(\TauCut) \Bigl[ \frac{\as}{2\pi} \bigl( c'_{1,1}  L_\Tau + c'_{1,0}\bigr) \Bigr]
   \nn\\&
%   + \cL''(\TauCut) \Bigl( - \frac{\zeta_2}{2}\Bigr)
   + \cL''(\TauCut) c''_{0,0} %\Bigl( - \frac{\zeta_2}{2}\Bigr)   
   + \cO(\as^3) \Bigr\}
   \,.\end{align}
%\commentgz{define $c''_{2,0}$?}
Here and in the following, we always use the abbreviations
\begin{align} \label{eq:muB_cumulant}
 \as = \as\Bigl(\sqrt{Q \TauCut}\Bigr) \,, \qquad L_\Tau = \frac12 \ln{\frac{Q}{\TauCut}}
\,.\end{align}
The constants appearing in \eq{TauCut_resummed_5} are given by
\begin{align} \label{eq:Tau0_constants}
 c_{1,0} &
 = A^{(1)} \zeta_2
\,,\nn\\
  c_{2,2} &
 = -8 [A^{(1)}]^2 \zeta_2
\,,\nn\\
 c_{2,1} &
 =  - 8 [A^{(1)}]^2 \zeta_3 + 4 A^{(1)} \zeta_2 \left(B_S^{(1)}-B_H^{(1)}\right) + 8 \pi \beta_0 A^{(1)} \zeta_2
 \nn\\&
 = - 4 A^{(1)} \bigl( c'_{1,0} - 2 \pi \beta_0 \zeta_2 \bigr)
\,,\nn\\
 c_{2,0} &
 = \zeta_2 A^{(2)} + \frac{3}{4} [A^{(1)}]^2 \zeta_4
 + 2 A^{(1)} \zeta_3 \left(B_S^{(1)}-B_H^{(1)}\right)
 - \frac{\zeta_2}{2} \left(B_S^{(1)}-B_H^{(1)}\right)^2
 \nn\\&\quad
 + \pi \beta_0 \bigl[ 4 A^{(1)} \zeta_3 +  \zeta_2 (B_H^{(1)}-3 B_S^{(1)}) \bigr]
 \nn\\&
 = A^{(2)} \zeta_2 + [A^{(1)}]^2 \left( \frac{3}{4} \zeta_4 + \frac{2 \zeta_3^2}{\zeta_2} \right)
 - \frac{(c'_{1,0})^2}{2 \zeta_2}
 + 2 \pi \beta_0 \Bigl( A^{(1)} \zeta_3 - B_S^{(1)} \zeta_2 + \frac12 c'_{1,0} \Bigr)\,,\nn\\
 c'_{1,0} &= 2 A^{(1)} \zeta_3  + \zeta_2 (B_H^{(1)} - B_S^{(1)})
\,,\nn\\
 c'_{1,1} &
 = 4 A^{(1)} \zeta_2
\,,\nn\\
  c''_{0,0} & = - \frac{\zeta_2}{2}
\,.\end{align}
Furthermore, we require the first and second derivative of $\cL$ up to $\alpha_s$ and $\alpha_s^2$, respectively.
They are given by
\begin{align} \label{eq:cL'}
 \cL'(\TauCut) &
 = \sum_{a,b} \frac{\df |\cM_{ab}|^2}{\df \PhiB}
   \bigl[
   - (\hat P \otimes f)_a \, f_{b}
   + (a \leftrightarrow b)
   \bigr]
\,,\\  \label{eq:cL''}
 \cL''(\TauCut) &
 = \sum_{a,b} \frac{\df |\cM_{ab}|^2}{\df \PhiB}
   \bigl[ \big\{(\hat P \otimes f)_a\, (\hat P \otimes f)_b
        - (\hat P' \otimes f)_a\, f_b
   \nn\\&\hspace{2.5cm}
        + (\hat P \otimes \hat P \otimes f)_a\, f_b\big\}
        + (a\leftrightarrow b) 
        \bigr]
\,,\end{align}
where the regularised splitting function $\hat P_{ij}$ can be expanded as 
%is defined as\commentgz{can be expanded as}\commentgz{coefficients needs indices...}
  \begin{align}
\label{eq:expP}
 \hat P_{ij}(x, \as) &= \sum_{n=0}^\infty \hat P^{(n)}_{ij}(x) \left(\frac{\as}{2\pi}\right)^{n+1}\,,
\end{align}
and its derivative is given by
\begin{align}
 \hat P_{ij}'(x,\as(\mu_B))
 = - \frac{\df}{\df\ln\TauCut} \hat P_{ij}\bigl(x, \as(\mu_B) \bigr)
 = \Bigl(\frac{\as(\mu_B)}{2\pi}\Bigr)^2 2 \pi \beta_0 \hat  P_{ij}^{(0)}(x) + \cO(\as^3)
\,.\end{align}

It is possible to absorb all terms in the first line of \eq{TauCut_resummed_5}
into a redefinition of the Sudakov and the luminosity. However, the $\as L_\Tau$ term in the second line of \eq{TauCut_resummed_5} can not be absorbed into a redefinition of the Sudakov without reintroducing such a term in the first line as well.
Similarly, the first term in \eq{cL''} can not be factorized such that it can be absorbed
into a redefinition of the beam function matching kernels $C_{ij}$. Therefore, we can express our final result as
\begin{align} \label{eq:MiNNLO_TauCut}
 \frac{\df \sigma^{\rm sing}(\TauCut)}{\df\PhiB} &
 = e^{-\cS(\TauCut)} \tilde\cL(\TauCut)
\,,\end{align}
by defining the final luminosity factor through
\begin{align} \label{eq:Tau0_cL_final}
 \tilde\cL(\TauCut) &
 = \sum_{a,b} \frac{\df |\cM_{ab}|^2}{\df \PhiB} \tilde H_{ab}(\mu_B)
   \Bigl[ (\tilde C \otimes f)_a \, (\tilde C \otimes f)_b
%     -\zeta_2 (\hat P \otimes f)_a\, (\hat P \otimes f)_b \Bigr]
     +2 c''_{0,0} (\hat P \otimes f)_a\, (\hat P \otimes f)_b \Bigr]   
 \nn\\&
 - \sum_{a,b} \frac{\df |\cM_{ab}|^2}{\df \PhiB}\Bigl(\frac{\as}{2\pi}\Bigr)^2 c'_{1,1} L_\Tau
   \Bigl[ (\hat P^{(0)} \otimes f)_a \, f_{b}
+ f_{a} \, (\hat P^{(0)} \otimes f)_b \Bigr] + \cO(\as^3)
\,.\end{align}
Here, the second line arises from the corresponding term in \eq{cL''},
and can not be absorbed into a redefinition of the kernels $C_{ij}$,
while the last line contains an explicit $L_\Tau$ term.
All non-logarithmic terms in \eq{TauCut_resummed_5} are absorbed
into the modified hard function and matching kernels, which are given by
\begin{align} \label{eq:def_tildeH}
 \tilde H_{ab}(\mu_B) &
 = H_{ab}(\mu_B) \biggl[ 1 + \frac{\as}{2\pi} c_{1,0} + \left(\frac{\as}{2\pi}\right)^2 c_{2,0} + \cO(\as^3) \biggr]
%%%
\,,\\ \nn
%%%
 \tilde C_{ij}(x, \mu_B) &
 = C_{ij}(x, \mu_B)
   - \Bigl(\frac{\as}{2\pi}\Bigr)^2 \left[
     - c''_{0,0} \bigl(\hat P^{(0)} \otimes \hat P^{(0)}\bigr)_{ij}(x)
%      \frac{\zeta_2}{2} \bigl(\hat P^{(0)} \otimes \hat P^{(0)}\bigr)_{ij}(x)     
      + \bigl( c'_{1,0} + 2 c''_{0,0} \pi \beta_0 \bigr) \hat P^{(0)}_{ij}(x)
%      + \bigl( c'_{1,0} - \zeta_2 \pi \beta_0 \bigr) \hat P^{(0)}_{ij}(x)      
   \right]
%  \nn\\&\quad
 + \cO(\as^3)
%  \tilde C_{ij}(x, \mu_B) &
%  = C_{ij}(x, \mu_B)
%    + \Bigl(\frac{\as}{2\pi}\Bigr)^2 \left[
%       - \frac{\zeta_2}{2} \bigl(P^{(0)} \otimes P^{(0)}\bigr)_{ij}(x)
%       - \bigl( c'_{1,1} L_\Tau + c'_{1,0} - \zeta_2 \pi \beta_0 \bigr) P^{(0)}_{ij}(x)
%    \right]
%  \nn\\&\quad
%  + \cO(\as^3)
 \,.\end{align}

In the original \minnlo{}-$\pt$ implementation, the remaining terms appearing
in the equivalent of \eq{TauCut_resummed_5} were absored into a
redefinition of the Sudakov form factor (specifically, by redefining
the $B$ coefficients). In our default
implementation of \minnlo{}-$\Tau_0$ we do not perform such a
redefinition.

Finally, we note that for the practical implementation it is convenient to combine
the two integrals in \eqn{eq:Tau0_L_S} into a single integral,
\begin{align}\label{eq:Tau0_Sud}
 \cS(\TauCut) &
 = 2 \int_0^{\ln\sqrt{Q/\TauCut}} \df \ell \Bigl\{ 2 A[\as(Q e^{-\ell})] \ell + 2 A[\as(\TauCut e^{\ell})] \ell
   \nn\\&\hspace{3cm}
   +  B_H[\as(Q e^{-\ell})] -  B_S[\as(\TauCut e^{\ell})] \Bigr\}
\,.\end{align}
For completeness, we also quote the expansion of the Sudakov,
\begin{align}
 S(\TauCut) & =
 \frac{\as(\mu_B)}{2\pi} \left[4 A^{(1)} L_\Tau^2 + 2 L_\Tau \bigl( B_H^{(1)}- B_S^{(1)} \bigr)\right]
 \nn\\&\quad
 + \left[\frac{\as(\mu_B)}{2\pi}\right]^2
   \left[ 4 L_\Tau^2 \left(A^{(2)} - \pi\beta_0 ( B_H^{(1)} +  B_S^{(1)}) \right)
          + 2 L_\Tau \bigl( B_H^{(2)}- B_S^{(2)}\bigr) \right]
 \nn\\&\quad
 + \left[\frac{\as(\mu_B)}{2\pi}\right]^3
   \biggl[ \frac{2}{3} (4 \pi \beta_0)^2 \bigl[  A^{(1)} L_\Tau^4 + L_\Tau^3 \bigl( B_H^{(1)}-  B_S^{(1)} \bigr)\bigr]
         \nn\\&\hspace{3cm}
         + L_\Tau^2 \left[ 4 A^{(3)} - 8 \pi^2 \beta_1 \bigl( B_H^{(1)} +  B_S^{(1)}\bigr) - 8 \pi \beta_0 \bigl(  B_H^{(2)} + B_S^{(2)}\bigr) \right]
         \nn\\&\hspace{3cm}
         + 2 L_\Tau \bigl( B_{H}^{(3)}- B_{S}^{(3)} \bigr)
   \biggr]
   + \cO(\as^4)\,.
 \end{align}

\subsection{Derivation of the \minnlo{} master formula for $\Tau_0$}
\label{sec:tau0master}

With the resummation formula in a suitable form we are now ready to derive the
 \minnlo{}-$\Tau_0$ master formula.
We start from the resummed cumulative cross section in \eqn{eq:TauCut_resummed_4},
\begin{align}\label{eq:cumulanttau}
 \frac{\df \sigma^{\rm sing}(\TauCut)}{\df\PhiB} &
 = \tilde \cL(\TauCut) e^{-\cS(\TauCut)}
\,,\end{align}
where all the ingredients have been derived in detail in the previous section.
We obtain the NNLO cross section differential in $\Tau_0$ 
 and in the Born phase space $\PhiB$ by taking the total derivative and then
 matching the ensuing differential 
  cross section at small $\Tau_0$ with the fixed-order cross section valid at large $\Tau_0$ 
\begin{align} \label{eq:start}
 \frac{\mathd\sigma}{\mathd\PhiB\mathd \Tau_0} &
  = \frac{\mathd}{\mathd \Tau_0} \bigg\{ \exp[-\cS(\Tau_0)] \, \tilde \cL(\Tau_0)\Bigg\} + R_f(\Tau_0)
 \nn\\&
  = \exp[-\cS(\Tau_0)]\left\{ D(\Tau_0)+\frac{R_f(\Tau_0)}{\exp[-\cS(\Tau_0)]}\right\}\,.
\end{align}
%\commentgz{e versus exp..}
Here, $R_f$ contains terms that are non-singular in the $\Tau_0\rightarrow 0$ limit, and we
have defined
\begin{equation}
\label{eq:Dterms}
  D(\Tau_0)  \equiv -\frac{\mathd \cS(\Tau_0)}{\mathd \Tau_0} \tilde\cL(\Tau_0)+\frac{\mathd \tilde\cL(\Tau_0)}{\mathd \Tau_0}\,.
\end{equation}
A central feature of the \minnlo{} method is that the renormalization
and factorization scales are evaluated at the low scale, i.e.\ the typical
scale of the resummation.\footnote{Note that this is not a strict requirement
of the \minnlo{} method and other choices are possible in general, but it has 
been introduced  as the default setting within the \minnlo{} approach so far.}
For $\Tau_0$ resummation there are two relevant low scales,
namely $\mu_S=\Tau_0$ and $\mu_B=\sqrt{\Tau_0\,Q}$, as discussed above.
We select $\muR\sim\muF\sim \mu_B$ as the common low scale
at which to evaluate all terms in \eq{start},
especially because the PDFs are naturally evaluated at that scale.
The logarithms to be resummed are then given by either
$\ln(Q/\mu_B) = \ln\sqrt{Q/\Tau_0}$ or $\ln(\Tau_0/\mu_B) = \ln\sqrt{\Tau_0/Q}$.
To simplify the notation in the following, we will always use
\begin{align} \label{eq:muB_spectrum}
 \mu_B = \sqrt{Q \Tau_0} = Q e^{-L_\Tau}
\,,\qquad
 L_\Tau = \frac12 \ln\frac{Q}{\Tau_0}
\,,\end{align}
which is the natural extension of \eq{muB_cumulant} to the $\Tau_0$ spectrum.
In particular, we express all derivatives w.r.t.~$\Tau_0$ as derivatives w.r.t.~$L_\Tau$.
This induces a Jacobian in \eq{Dterms},
\begin{equation} \label{eq:Dterms_2}
 D(\Tau_0)
 = \frac{1}{2 \Tau_0} \left[ \frac{\df \cS(\Tau_0)}{\df L_\Tau} \tilde\cL(\Tau_0) - \frac{\df \tilde\cL(\Tau_0)}{\df L_\Tau} \right]
\,,\end{equation}
but simplifies the expressions for derivatives of $\cS$ and $\tilde\cL$ evaluated below.

As in \eqn{eq:NLO} for \minnlo{}-$\pt$, we 
introduce the NLO differential cross section for \FJ{} production
\begin{equation}
\label{eq:NLOtau}
\frac{\mathd\sigma^{\rm (NLO)}_{\scriptscriptstyle\rm FJ}}{\mathd\PhiB\mathd
      \Tau_0} = \abarmu{\mu_B}\left[\frac{\mathd\sigma_{\scriptscriptstyle\rm FJ}}{\mathd\PhiB\mathd
      \Tau_0}\right]^{(1)} + \left(\abarmu{\mu_B}\right)^2\left[\frac{\mathd\sigma_{\scriptscriptstyle\rm FJ}}{\mathd\PhiB\mathd
      \Tau_0}\right]^{(2)}\,,
\end{equation}
so that we can write the finite remainder $R_f$ at this order as
\begin{align}\label{eq:R_f}
R_f = \frac{\mathd\sigma^{\rm (NLO)}_{\scriptscriptstyle\rm FJ}}{\mathd\PhiB\mathd
      \Tau_0} - \abarmu{\mu_B} \left[\exp[-\cS(\Tau_0)] D(\Tau_0)\right]^{(1)} - \left(\abarmu{\mu_B}\right)^2 \left[\exp[-\cS(\Tau_0)] D(\Tau_0)\right]^{(2)}\,.
\end{align}
%\begin{align} \label{eq:R_f}
% R_f(\Tau_0)
% = \frac{\mathd\sigma_{\scriptscriptstyle\rm FJ}}{\mathd\PhiB\mathd\Tau_0} \bigg|_{\rm f.o.}
% - \frac{\df \sigma^{\rm sing}}{\df\PhiB \df \TauCut}
% = \frac{\mathd\sigma_{\scriptscriptstyle\rm FJ}}{\mathd\PhiB\mathd\Tau_0} - \exp[-\tilde \cS(\Tau_0)] D(\Tau_0) \bigg|_{\rm f.o.}
%\,,\end{align}
%where we make explicit that the $R_f$ is always expanded at fixed-order,
%as opposed to the $D$ term in \eq{start} that is kept exact.
This allows us to rewrite \eq{start} as
\begin{align}
\label{eq:minnlo-tau}
  \frac{\mathd\sigma}{\mathd\PhiB\mathd \Tau_0}
  & =  \exp[-\cS(\Tau_0)] \bigg\{
      \abarmu{\mu_B} \left[\frac{\mathd\sigma_{\scriptscriptstyle\rm FJ}}{\mathd\PhiB\mathd\Tau_0}\right]^{(1)} \left(1+\abarmu{\mu_B} [\cS(\Tau_0)]^{(1)}\right)
    \nn\\&
    + \left(\abarmu{\mu_B}\right)^2\left[\frac{\mathd\sigma_{\scriptscriptstyle\rm FJ}}{\mathd\PhiB\mathd \Tau_0}\right]^{(2)}
    \nn
    + \left[D(\Tau_0) -\abarmu{\mu_B} [D(\Tau_0)]^{(1)} -\left(\abarmu{\mu_B}\right)^2 [D(\Tau_0)]^{(2)}  \right] 
   \nn\\&
   + {\rm regular~terms~of~{\cal O}(\as^3)}
 \bigg\}
\,.\end{align}
Here, $D(\Tau_0)$ is the term from \eq{start} without any further expansion,
while the other terms arise from expanding $R_f/e^{-\cS}$ using \eq{R_f}.\footnote{
Note that, when deriving $D(\Tau_0)$ all resummation ingredients need to be included to
an order such that the cumulative cross section in \eqn{eq:cumulanttau} is accurate at relative
$\mathcal{O}(\alpha_s^2)$, in particular that means two-loop order in the 
hard function $H$, the collinear kernels $C_{ij}$, as well as the beam and soft function.
This is necessary to ensure NNLO accuracy of the \minnlo{} approach.}
Using \eq{Dterms_2}, the $[D(\Tau_0)]^{(n)}$ coefficients are given by
\begin{align}
 [D(\Tau_0)]^{(1)} &
 = \frac{1}{2 \Tau_0} \Biggl\{ \biggl[\frac{\mathd \cS(\Tau_0)}{\df L_\Tau}\biggr]^{(1)}[\tilde\cL(\Tau_0)]^{(0)}
                              - \biggl[\frac{\mathd \tilde\cL(\Tau_0)}{\df L_\Tau}\biggr]^{(1)} \Biggr\}
\,,\\\nn
 [D(\Tau_0)]^{(2)} &
 = \frac{1}{2 \Tau_0} \Biggl\{ \biggl[\frac{\mathd \cS(\Tau_0)}{\df L_\Tau}\biggr]^{(2)} [\tilde\cL(\Tau_0)]^{(0)}
                             + \biggl[\frac{\mathd \cS(\Tau_0)}{\df L_\Tau}\biggr]^{(1)}[{\tilde\cL}(\Tau_0)]^{(1)}
                             - \biggl[\frac{\mathd \tilde\cL(\Tau_0)}{\df L_\Tau}\biggr]^{(2)} \Biggr\}
\,.\end{align}
The expansion of the derivative of the Sudakov form factor with respect to $L_\Tau = \ln\sqrt{Q/\TauCut}$  are given by
\begin{align}
  \label{eq:S_derivative}
\left[ \frac{\df S}{\df L_\Tau}\right]^{(1)} &
 =
 %\frac{\as(\mu_B)}{2\pi}
 2 \bigl( 4 A^{(1)} L_\Tau+ B_H^{(1)}- B_S^{(1)}\bigr)
 \nn\,,\\\quad
 \left[\frac{\df S}{\df L_\Tau}\right]^{(2)} & =
 %\left[
   %\frac{\as(\mu_B)}{2\pi}
   %\right]^2
   2 \left[ 4 A^{(2)} L_\Tau+ B_H^{(2)}- B_S^{(2)}
          + 8 \pi \beta_0 \bigl(A^{(1)} L_\Tau^2 - B_S^{(1)}L_\Tau \bigr) \right]\,,
\end{align}
%are given in \eq{S_derivative},
while the fixed-order terms of $\tilde\cL(\Tau_0)$ and their derivatives 
follow from \eq{Tau0_cL_final},
\begin{align} \label{eq:Tau0_ingredients_2}
 [\tilde\cL(\Tau_0)]^{(0)} &
 = \sum_{a,b} \frac{\df |\cM_{ab}|^2}{\df \PhiB} f_{a} f_{b}
%%%
\,,\\\nn
%%%
 [\tilde\cL(\Tau_0)]^{(1)} &
 = \sum_{a,b} \frac{\df |\cM_{ab}|^2}{\df \PhiB}
   \Bigl\{ \tilde H^{(1)} f_a f_b + (\tilde C^{(1)} \otimes f)_a \, f_b +  f_a \, (\tilde C^{(1)} \otimes f)_b \Bigr\}
%  \\\nn&
%  = \sum_{c,c'}\frac{\mathd|M^{\scriptscriptstyle\rm F}|_{cc'}^2}{\mathd\PhiB}\bigg\{H^{(1)}f_c^{[a]}f_{c'}^{[b]} + (C^{(1)}\otimes f)_c^{[a]}f_{c'}^{[b]} + f_{c}^{[a]} (C^{(1)}\otimes f)_{c'}^{[b]}\bigg\}\,,  CORRECT?!
%%%
\,,\\\nn
%%%
 [\tilde\cL(\Tau_0)]^{(2)} &
 = \sum_{a,b} \frac{\df |\cM_{ab}|^2}{\df \PhiB}
   \Bigl\{
   \tilde H^{(1)} \Bigl[ (\tilde C^{(1)} \otimes f )_a \, f_b + f_a \,(\tilde C^{(1)} \otimes f)_b \Bigr]
   + \tilde H^{(2)} f_a f_b
   \\\nn&\hspace{2.8cm}
   + (\tilde C^{(2)} \otimes f)_a\, f_b + f_a\, (\tilde C^{(2)} \otimes f)_b
   \\\nn&\hspace{2.8cm}
   + ( \tilde C^{(1)} \otimes f)_a\, ( \tilde C^{(1)} \otimes f)_b
   - \zeta_2 ( \hat P^{(0)} \otimes f)_a\, ( \hat P^{(0)} \otimes f)_b
   \\\nn&\hspace{2.8cm}
   - c'_{1,1} L_\Tau \Bigl[(\hat  P^{(0)} \otimes f)_a \, f_b + f_a \, (\hat P^{(0)} \otimes f)_b \Bigr]
   \Bigr\}
%%%
\,,\\\nn
%%%
 \left[ \frac{\df\tilde\cL(\Tau_0)}{\df L_\Tau} \right]^{(1)} &
 = \sum_{a,b} \frac{\df |\cM_{ab}|^2}{\df \PhiB} (-2)
   \Bigl\{ (\hat P^{(0)} \otimes f)_a\, f_b + f_a \, (\hat P^{(0)} \otimes f)_b \Bigr\}
% \,,\\\nn
% \left[  \frac{\mathd \tilde\cL(\Tau_0)}{\mathd \Tau_0}
%   \right]^{(1)} & = \sum_{c,c'}\frac{\mathd|M^{\scriptscriptstyle\rm F}|_{cc'}^2}{\mathd\PhiB}\frac{2}{\Tau_0}\bigg\{ (\hat{P}^{(0)}\otimes
%                              f)_c^{[a]}f_{c'}^{[b]} +  f_{c}^{[a]}
%                              (\hat{P}^{(0)}\otimes f)_{c'}^{[b]}\bigg\}\,, CORRECT?!
%%%
\,,\\\nn
%%%
 \left[ \frac{\df\tilde\cL(\Tau_0)}{\df L_\Tau} \right]^{(2)} &
 = \sum_{a,b} \frac{\df |\cM_{ab}|^2}{\df \PhiB} (-2)
   \Bigl\{
   (\hat P^{(1)} \otimes f)_a\, f_b + f_a \, (\hat P^{(1)} \otimes f)_b
   \\\nn&\hspace{3.5cm}
   + \Bigl( \tilde H^{(1)} + \frac{c'_{1,1}}{2}\Bigr) \Bigl[ (\hat P^{(0)} \otimes f\bigr)_a\, f_b + f_a\, (\hat P^{(0)} \otimes f)_b \Bigr]
   \\\nn&\hspace{3.5cm}
   + (\hat P^{(0)} \otimes f)_a\, (\tilde C^{(1)} \otimes f)_b
   + (\tilde C^{(1)} \otimes f)_a\, (\hat P^{(0)} \otimes f)_b
   \\\nn&\hspace{3.5cm}
   + (\tilde C^{(1)} \otimes \hat P^{(0)} \otimes f)_a\, f_b
   + f_a\, (\tilde C^{(1)} \otimes \hat P^{(0)} \otimes f)_b
   \\\nn&\hspace{3.5cm}
   -2 \pi \beta_0 \Bigl[ \tilde H^{(1)} f_a f_b + ( C^{(1)} \otimes f)_a\, f_b + f_a \, ( C^{(1)} \otimes f)_b \Bigr]
   \Bigr\}
\,.\end{align}
The scale dependence is implemented as in Appendix\,D of
\citere{Monni:2019whf}, which directly generalizes from the $\pt$ to the 
$\Tau_0$ case up to the required accuracy, as we have made sure that our starting equation has the 
identical structure.
We note that 
$[D(\Tau_0)]^{(1)} $ does not depend on the renormalization and
factorization scale factors $\KR$ and $\KF$, while for 
$ [D(\Tau_0)]^{(2)} $ one has 
\begin{align}
[D(\Tau_0)]^{(2)}(\KF,\KR) &= [D(\Tau_0)]^{(2)} - 2\beta_0 \pi \left[  \frac{\mathd \tilde\cL(\Tau_0)}{\mathd \Tau_0}
  \right]^{(1)} \ln\frac{\KF^2}{\KR^2}\,.
\end{align}

By construction, \eq{minnlo-tau} yields the NNLO
cross section fully differential in $\PhiB$ when integrating over $\Tau_0$, which can be
understood as follows:
Our starting equation \eq{start} is computed at exactly that accuracy,
provided that we take into account all relevant contributions up to two-loop level in the 
computation of the cumulant in \eqn{eq:cumulanttau}. Given that we have 
not expanded $D(\Tau_0)$ in \eq{minnlo-tau} the total derivative w.r.t.\ $\Tau_0$ in
\eqn{eq:start}, and therefore in $D(\Tau_0)$, is kept intact. As a result,
\eq{minnlo-tau} retains NNLO accuracy in $\PhiB$ (when integrating over $\Tau_0$).
By contrast, if we were to truncate $D(\Tau_0)$ beyond $\cO(\as^3)$, for instance,
we would neglect large logarithmic terms proportional to $L_\Tau^n$ at higher orders. 
Depending on their power $n$, these missing logarithmic terms could spoil the NNLO accuracy 
upon integration over $\Tau_0$. In the case of \minnlo{}-$\pt{}$ expanding $D(\pt{})$ 
up to $\cO(\as^3)$ would be sufficient to retain NNLO accuracy, but in the case
of $\Tau_0$ higher powers $n$ of the logarithms appear, which would 
require an expansion even beyond $\cO(\as^3)$. This renders it crucial to keep 
the full unexpanded $D(\Tau_0)$ for \minnlo{}-$\Tau_0$.

We can now apply
the \minnlo{}-$\Tau_0$ procedure directly at the fully differential level
in the $\PhiBJ$ phase space to the \POWHEG{} $\rm FJ{}$ calculation, as given in \eqn{eq:master}, 
by replacing the corresponding ${\bar B}(\PhiBJ)$ function through
\begin{align}
\label{eq:BbarTau0}
{\bar B}(\PhiBJ)&\equiv \exp[-\mathcal{S}(\Tau_0)]\bigg\{ \abarmu{\Tau_0}\left[\frac{\mathd\sigma_{\scriptscriptstyle\rm FJ}}{\mathd\PhiBJ}\right]^{(1)} \left(1+\abarmu{\Tau_0} [\tilde{S}(\Tau_0)]^{(1)}\right)\notag
  \\
&+ \left(\abarmu{\Tau_0}\right)^2\left[\frac{\mathd\sigma_{\scriptscriptstyle\rm FJ}}{\mathd\PhiBJ}\right]^{(2)} + D^{(\ge 3)}(\Tau_0)\,  F^{\tmop{corr}}(\PhiBJ)\bigg\}\,.
\end{align}
The factor $F^{\tmop{corr}}(\PhiBJ)$ encodes a suitable function to spread the correction $D^{(\ge 3)}(\Tau_0)$,
which intrinsically depends only $\Tau_0$ and $\PhiB$, on the full $\PhiBJ$ phase space, as discussed 
in detail in appendix \ref{app:spreading}.

\subsection{Matching with the shower}
\label{sec:truncshower}

So far, we presented how to reach NNLO accuracy within the \minnlo{} method using $\Tau_0$ as a resolution variable. We now discuss
  how to match our \minnlo{}-$\Tau_0$ predictions with a parton shower. In both \minnlo{}-$\pt$ and \minnlo{}-$\Tau_0$, the matching with the parton
  shower relies on the \POWHEG{} formalism. In particular, the non-emission probability associated to the first and to the second emissions
  are encoded in the \minnlo{} and \POWHEG{} Sudakov form factors, given in \eqn{eq:Tau0_Sud} for \minnlo{}-$\Tau_0$ (eq. (2.9) of~\cite{Monni:2019whf}
  for \minnlo{}-$\pt$) and \eqn{eq:master}, respectively.
  In the original \minnlo{}-$\pt$ formulation, these Sudakov
  form factors are both associated to transverse-momentum like observables, and they match the leading-logarithmic structure of
  a transverse-momentum ordered parton shower. This implies that, as long as the emissions generated by the shower are ordered in transverse momentum and vetoed according to the \POWHEG{} procedure,
  the leading-logarithmic accuracy of the shower is preserved. In this case, the corresponding Lund plane~\cite{Andersson:1988gp} is filled without leaving any empty areas and without covering the same
  area twice, which would constitute a leading-logarithmic violation. By contrast, in the \minnlo{}-$\Tau_0$ approach the \minnlo{} Sudakov form factor is
  associated to a different variable, $\Tau_0$, whose resummation has a different leading-logarithmic structure. Thus, the matching with a parton shower becomes a delicate point. Filling correctly the Lund plane is
  now highly non trivial, since applying the standard \POWHEG{} formalism can lead to empty areas or regions that are accounted for two times (thus wrongly suppressed).

\begin{figure}[t]
\begin{center}\vspace{-0.2cm}
\begin{tabular}{cc}
\includegraphics[width=.32\textheight,page=8]{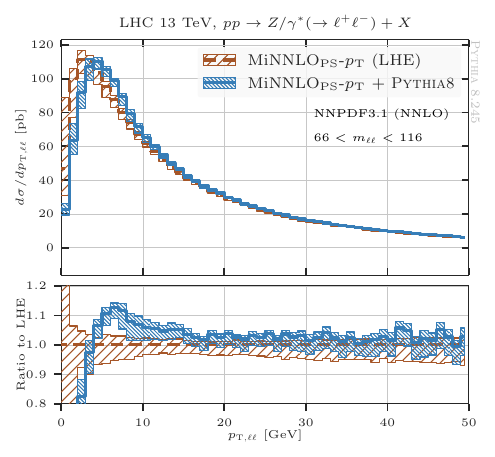}
&
\includegraphics[width=.32\textheight,page=8]{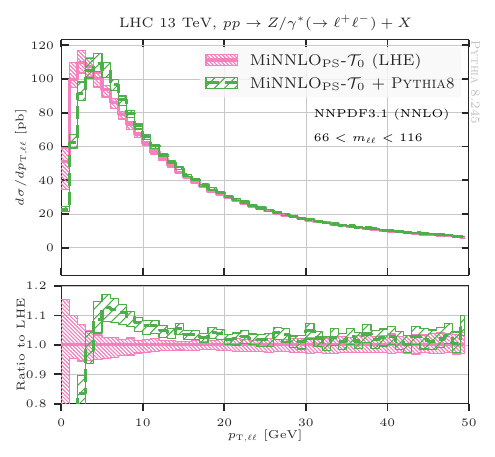}
\end{tabular}
\caption{\label{fig:lhevspy8} Transverse-momentum distribution of the lepton pair in Drell-Yan production. 
Left plot: comparison of \minnlo-$\pt$ at LHE level (brown, dashed) and after parton showering (blue, solid); 
right plot: comparison of  \minnlo-$\Tau_0$ at LHE level (pink, solid) and after parton showering (green, dashed).
The ratio plots show the relative effect of the parton shower.}
\end{center}
\end{figure}

  To address these issues, one should modify the \POWHEG{} mapping and/or rely on a truncated-vetoed shower~\cite{Nason:2004rx} to properly fill the Lund plane. More concretely, this implies a substantial modification of the \POWHEG{} code.
  We note that analogue modifications would be needed to implement a  matching with NLL-accurate parton showers, which are being actively developed at the time of writing~\cite{Bewick:2019rbu,Dasgupta:2020fwr,Forshaw:2020wrq,Nagy:2020rmk,Bewick:2021nhc,Herren:2022jej,vanBeekveld:2022zhl,vanBeekveld:2022ukn}.
  Since the main goal of this paper is to define the theoretical framework of the new \minnlo{} formalism based on jettiness-like observables, we prefer to postpone a detailed discussion on the matching to a future paper. We believe that studying this problem in a
  separate work would be optimal, in view of the construction of an NNLO Monte Carlo event generator consistently matched to an NLL parton shower.
  Therefore, in this paper we do not implement these modifications, which means that in our matched results the accuracy of
  the parton shower is not fully preserved. However, we will show that the numerical effect of this choice is rather small by comparing our \minnlo{}-$\Tau_0$ and \minnlo{}-$\pt$ results
  after showering. 
  This comparison provides us with an estimate of the size of the neglected terms needed to formally preserve the shower accuracy.  
  Moreover, the impact of the parton shower is almost identical for \minnlo{}-$\pt$ and \minnlo{}-$\Tau_0$.
  This can be observed from the two ratio panels of  \fig{fig:lhevspy8} that show the showered result divided by the Les-Houches-Event (LHE) level one
   for the example case of the transverse momentum distribution of the $Z$ boson in Drell-Yan production. We find an analogous behaviour for all distributions
   we considered in Higgs-boson and Drell-Yan production. 
   In conclusion, despite the fact that the logarithmic accuracy of the parton shower is not formally preserved in the \minnlo{}-$\Tau_0$ case, its effect in the two \minnlo{} formulations is almost identical.

\section{Validation and phenomenological results}
\label{sec:pheno}
In this section we present phenomenological results for 
Drell-Yan production
($pp \rightarrow \ell^+ \ell^-$) and on-shell Higgs-boson production ($pp \rightarrow H$) in the heavy-top limit.
For our practical implementation we use as a starting point the \minnlo{}-$\pt$ generators in 
\noun{POWHEG-BOX-V2} developed in \citeres{Monni:2019whf,Monni:2020nks}, which are based on the \POWHEG{} computations of $H$+jet~\cite{Campbell:2012am} and $Z$+jet~\cite{Alioli:2010qp} production, and we apply the \minnlo{}-$\Tau_0$ formalism discussed in the previous section.
The relevant input parameters are discussed in \sct{sec:setup}. First, we validate 
our \minnlo{}-$\Tau_0$ predictions against the \minnlo{}-$\pt$ ones~\cite{Monni:2019whf,Monni:2020nks} in \sct{sec:tau0vspt}.
Then, we present the first comparison between results from the \minnlo{} and {\sc Geneva} \cite{Alioli:2021qbf}
generators in \sct{sec:MiNNLOvsGeneva}. Finally, we compare our predictions
against high-precision data from ATLAS \cite{ATLAS:2019zci} and CMS \cite{CMS:2019raw} for Drell-Yan
production at 13\,TeV in \sct{sec:data}.

\subsection{Setup}
%\subsection{Input parameters and setup}
\label{sec:setup}

We consider proton--proton collisions at the LHC with a center-of-mass
energy of 13\,TeV. We use the $G_\mu $ scheme with $\cos^2 \theta_W = m_W^2/m_Z^2$ and $\alpha = \sqrt{2} G_\mu m_W^2 \sin^2 \theta^2_W/\pi$ and we set the electroweak (EW) inputs to their PDG~\cite{ParticleDataGroup:2020ssz} values:
$G_F = 1.16639 \times 10^{-5}$~GeV$^{-2}$, $m_W = 80.385$~GeV,
$\Gamma_W = 2.0854$~GeV, $m_Z = 91.1876$~GeV, $\Gamma_Z = 2.4952$~GeV,
$m_H = 125$~GeV.
We set the on-shell top-quark mass to $m_t = 173.2$~GeV.
Our choice for the parton densities is the NNLO set of
NNPDF3.1~\cite{NNPDF:2017mvq} with $\as=0.118$, which is obtained 
via the \textsc{lhapdf} interface~\cite{Buckley:2014ana}.
%for all our predictions.
The PDFs are read by
\textsc{lhapdf} and evolved internally through
\textsc{hoppet}~\cite{Salam:2008qg} as described in
\citere{Monni:2019whf}.  The central factorization and renormalization
scales are set following the \minnlo{} procedure, as described before.
For Higgs-boson production the overall two powers of the strong coupling 
are evaluated at the scale $\mu_R^{(0)}=m_H$.
We estimate the uncertainties due to missing higher-order corrections
through the usual variations of $\muF$ and $\muR$ around their central value by a factor of
two in each direction with the constraint $0.5 \leq \muR /\muF \leq 2$ while keeping the minimal and maximal values
of the cross section.
Resummation effects at large $\mathcal T_0$ are switched off by replacing the nominal logarithm $L_{\mathcal T}$ with the modified logarithm $\tilde L_{\mathcal T} =1/p \ln (1 + (\sqrt{Q/\mathcal T})^p)$.  For our predictions, we set $p=6$.
%\renewcommand{\baselinestretch}{1.5}
%\begin{table}[b]
%\centering
%  \begin{tabular}{l|cc}
%     & \setupinclusive & \setupfiducial \\
%    \toprule
%              $Z$-mass window & $60$\,GeV$ < m_{Z_1},m_{Z_2} < 120$\,GeV & $60$\,GeV$ < m_{Z_1},m_{Z_2}< 120$\,GeV\\[0.1cm]
%    lepton cuts & $m_{\ell^+\ell^-} > 4 \, {\rm GeV}$  & $\begin{array}{c}
%   p_{T,\ell_1} > 20\,{\rm GeV}, \quad p_{T,\ell_2} > 10 \, {\rm GeV},  \\[-0.15cm]
%   p_{T,\ell_{3,4}} > 5 \, {\rm GeV}, \quad |\eta_\ell | < 2.5,\\[-0.15cm]
%   m_{\ell^+\ell^-} > 4 \, {\rm GeV}  \\    
% \end{array}$
% \end{tabular}
% \renewcommand{\baselinestretch}{1.0}
%  \caption{Inclusive and fiducial cuts used to the define the \setupinclusive and \setupfiducial phase space regions~\cite{CMS:2020gtj}. See text for more details.}
%   \label{tab:cuts}
%\end{table}
% \renewcommand{\baselinestretch}{1.0}
For all predictions presented in this paper we make use of the \PYTHIA{8}
parton shower~\cite{Sjostrand:2014zea} and we employ a variation of the 
{\sc Monash} tune~\cite{Skands:2014pea} adapted by CMS to improve the 
description of the Drell-Yan transverse-momentum spectrum.\footnote{We thank Kenneth Long 
for providing us with the settings.}

We validate our \minnlo{}-$\Tau_0$ implementation for Drell-Yan and Higgs production 
against reference predictions from the \minnlo{}-$\pt{}$ generators developed 
in \citeres{Monni:2019whf,Monni:2020nks} using the identical input settings.
Moreover, we present a first comparison between \minnlo{} and {\sc Geneva}
predictions for Drell-Yan production. The {\sc Geneva} results 
correspond to those presented in \citere{Alioli:2021qbf}, and we refer the reader to that 
paper for the respective input parameters and settings.
Finally, we compare our predictions with experimental measurements by both 
the ATLAS \cite{ATLAS:2019zci} and the CMS collaboration~\cite{CMS:2019raw}.
In order to provide the most realistic comparison to experimental
data, our showered predictions always include effects from hadronization and
multi-particle interactions (MPI). We do not require any lepton dressing, as we do not include 
any QED showering effects.

\subsection{Comparison and validation against \minnlo{}-$\pt$ results}
\label{sec:tau0vspt}
We start the presentation of the phenomenological results by comparing 
our \minnlo{}-$\Tau_0$ predictions with \minnlo{}-$\pt{}$ ones. The \minnlo{}-$\pt{}$
generators have been tested extensively against fixed-order predictions in \citere{Monni:2020nks}.
They therefore serve us as reference predictions to validate our \minnlo{}-$\Tau_0$ implementations
in this section.

\subsubsection{Total cross section}

\renewcommand\arraystretch{1.3}
\begin{table}[ht]
\begin{center}
%\resizebox{\columnwidth}{!}{%
\begin{tabular}{c l c | c c c c c c c}
\toprule
%\multirow{2}{*}{inclusive}
 &&&& \multicolumn{2}{c}{$pp\to H$ (on-shell)}
&& \multicolumn{2}{c}{$pp\to Z \to \ell^+\ell^-$} &\\
 &&&& $\sigma$ [pb]
& $\sigma/\sigma_{\rm NNLO}$
&& $\sigma$ [fb]
& $\sigma/\sigma_{\rm NNLO}$ &\\
\midrule
&NNLO  &&& $40.32(2)_{-10.4\%}^{+10.7\%}$ & 1.000 && $1919(1)_{-1.1\%\phantom{1}}^{+0.9\%\phantom{1}}$ & 1.000&\\
%&\minlo{}  &&& $30.40(3)_{-15.0\%}^{+33.3\%}$ & 0.767 && $1837(2)_{-14.5\%}^{+20.1\%}$ & 0.943 &\\
&\minnlo{}-$\pt{}$  &&& $39.33(1)_{-11.0\%}^{+12.2\%}$ & 0.975 && $1907(2)_{-1.2\%\phantom{1}}^{+1.1\%\phantom{1}}$ & 0.994 &\\
&\minnlo{}-$\Tau_0$  &&& $41.56(2)_{-10.1\%}^{+9.4\%}$ & 1.031 && $1925(1)_{-1.2\%\phantom{1}}^{+1.2\%\phantom{1}}$ & 1.003 &\\
\bottomrule
\end{tabular}%}
\end{center}
\renewcommand{\baselinestretch}{1.0}
\caption{ \label{tab:cs} Predictions of the total inclusive cross section for Higgs-boson 
production and the DY process at NNLO obtained with \Matrix{} \cite{Grazzini:2017mhc}, and using the 
\minnlo{}-$\pt$ and \minnlo{}-$\Tau_0$ implementations.
The second and fourth columns show the ratio to the NNLO cross section.
}
\end{table}
\renewcommand\arraystretch{1}

\Tab{tab:cs} compares \minnlo{}-$\Tau_0$ and \minnlo{}-$\pt{}$ predictions for the total inclusive cross sections 
for Higgs-boson production and for Drell-Yan production (with an invariant-mass window of $66\,{\rm GeV} < \mll < 116\,{\rm GeV}$).
One should bear in mind that, despite both being NNLO accurate, these predictions  differ by terms beyond NNLO accuracy.
This is the case as they use different matching observables in their expansion and different scale settings. As a result, those predictions
should agree within the quoted perturbative uncertainties.
Indeed, as one can see, the predicted rates from \minnlo{}-$\Tau_0$ and \minnlo{}-$\pt{}$ are fully consistent with each other within
the uncertainties from scale variation.

\subsubsection{NNLO accuracy in distributions of the colour-singlet final states}

\begin{figure}[t]
\begin{center}\vspace{-0.2cm}
\includegraphics[width=.45\textheight, page=1]{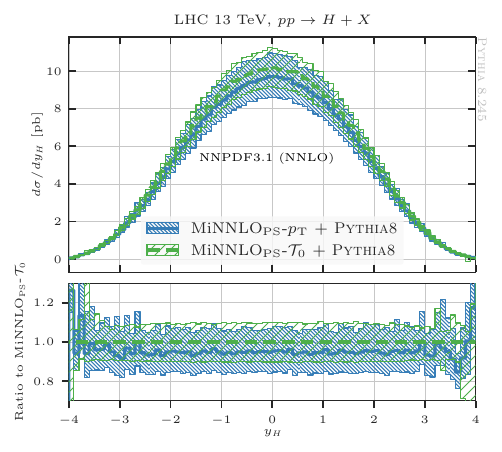}
\caption{\label{fig:Higgsrapidity} Comparison of \minnlo{}-$\Tau_0$ (green, dashed) and \minnlo{}-$\pt{}$ (blue, solid) predictions for the rapidity distribution of the Higgs boson.}
\end{center}
\end{figure}

%\afterpage{
\begin{figure}[t]
\begin{center}\vspace{-0.2cm}
\begin{tabular}{cc}
\includegraphics[width=.32\textheight, page=1]{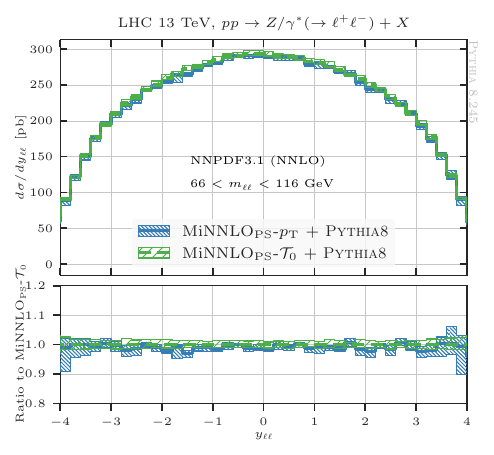}
&
\includegraphics[width=.32\textheight, page=2]{plots/MiNNLO_Tau0_pT_dists.pdf}
\end{tabular}\vspace{0.3cm}
\begin{tabular}{cc}
\includegraphics[width=.32\textheight, page=5]{plots/MiNNLO_Tau0_pT_dists.pdf}
&
\includegraphics[width=.32\textheight, page=3]{plots/MiNNLO_Tau0_pT_dists.pdf}
\end{tabular}
%\vspace{0.3cm}
%\begin{tabular}{cc}
%\includegraphics[width=.32\textheight, page=6]{plots/MiNNLO_Tau0_pT_dists.pdf}
%&
%\includegraphics[width=.32\textheight, page=4]{plots/MiNNLO_Tau0_pT_dists.pdf}
%\end{tabular}
\caption{\label{fig:DYvalidation} Comparison of \minnlo{}-$\Tau_0$ (green, dashed) and \minnlo{}-$\pt{}$ (blue, solid) predictions for differential distributions in the phase-space of the leptons in Drell-Yan production.}
\end{center}
\end{figure}
%\clearpage}

We continue by considering differential distributions of the colour-singlet final states in Higgs-boson and Drell-Yan production.
Since the results at Les Houches event (LHE) level are similar to the showered ones for all the distributions discussed in this section, we only show results including shower effects as well as hadronisation and multi-parton interactions (MPI).
For on-shell Higgs-boson production the only formally NNLO-accurate observable is the rapidity of the Higgs boson, whose 
distribution is shown in \fig{fig:Higgsrapidity}. The predictions from the \minnlo{}-$\Tau_0$ (green, dashed) and \minnlo{}-$\pt{}$ (blue, solid) 
generators are 
in complete agreement within the given uncertainty bands. This is a numerical confirmation of the NNLO accuracy of 
our new \minnlo{}-$\Tau_0$ implementation for Higgs boson production.
Similarly, we present a validation of NNLO-accurate observables in the phase space of the final-state leptons in Drell-Yan production
in \fig{fig:DYvalidation}. In particular, the distributions in the rapidity ($\yll$) and invariant-mass ($\mll$) of the lepton pair, 
as well as the rapidity ($\ylp$) and the transverse-momentum ($\ptlp$) of the positively charged lepton are shown. In all cases, we find a remarkable agreement
between the \minnlo{}-$\Tau_0$ and \minnlo{}-$\pt{}$ results, especially considering the very small scale uncertainties of this process, which are
of order $\sim 1\%$. Only for $\ptlp\gtrsim m_Z/2$ the uncertainty bands increase to 5-10\% and the two predictions differ by about 5\% from each other 
(i.e.\ again within those uncertainties). This behaviour is well understood and can be traced back to a phase-space effect~\cite{Catani:1997xc}, which requires the two (back-to-back) leptons at LO to share the available energy of $\mll\sim m_Z$ among them, effectively restricting their transverse momentum spectra
to $\ptl\lesssim m_Z/2$. As a result, the $\ptlp$ distribution, even in an NNLO calculation, becomes formally only NLO accurate above the $m_Z/2$
 threshold, which explains both the increased uncertainty band and the larger differences between the \minnlo{}-$\Tau_0$ and \minnlo{}-$\pt{}$ predictions.

%\begin{figure}[t]
%\begin{center}\vspace{-0.2cm}
%\includegraphics[width=.45\textheight, page=9]{plots/MiNNLO_Tau0_pT_dists.pdf}
%%\begin{tabular}{cc}
%%\includegraphics[width=.32\textheight, page=7]{plots/MiNNLO_Tau0_pT_dists.pdf}
%%&
%%\includegraphics[width=.32\textheight, page=8]{plots/MiNNLO_Tau0_pT_dists.pdf}
%%\end{tabular}\vspace{0.3cm}
%\caption{\label{fig:data1} ... remove if we compare MiNNLO tau0 and pT to data, if only tau0 compared to data keep?}
%\end{center}
%\end{figure}

\subsubsection{NLO accuracy in exclusive distributions in the one-jet phase space}

\afterpage{
\begin{figure}[t]
\begin{center}\vspace{-0.2cm}
\begin{tabular}{cc}
\includegraphics[width=.32\textheight, page=4]{plots/MiNNLO_Tau0_pT_dists_HIggs.pdf}
&
\includegraphics[width=.32\textheight, page=10]{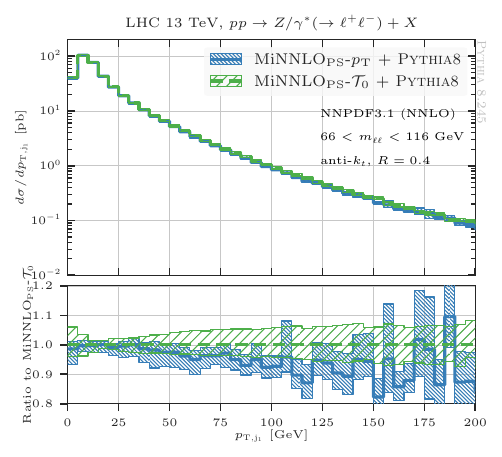}
\end{tabular}\vspace{0.3cm}
\begin{tabular}{cc}
\includegraphics[width=.32\textheight, page=5]{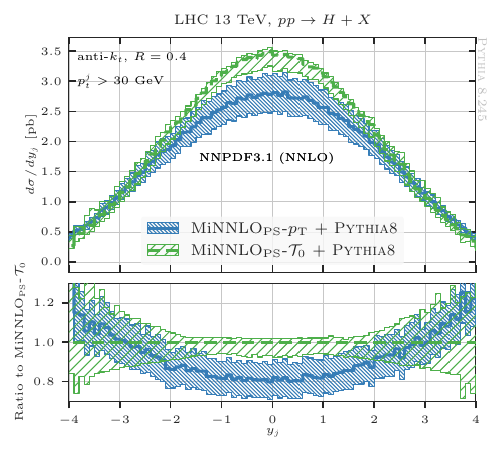}
&
\includegraphics[width=.32\textheight, page=11]{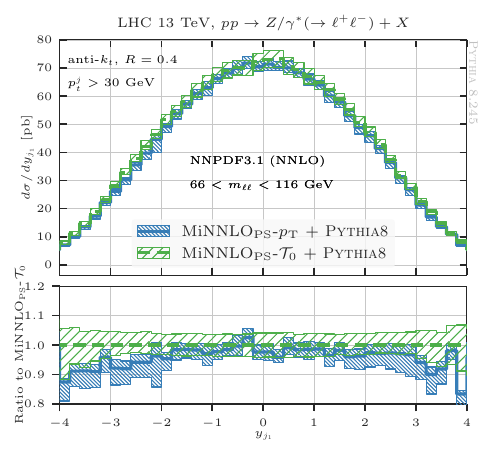}
\end{tabular}\vspace{0.3cm}
%\begin{tabular}{cc}
%\includegraphics[width=.32\textheight, page=12]{plots/MiNNLO_Tau0_pT_dists.pdf}
%&
%\includegraphics[width=.32\textheight, page=12]{plots/MiNNLO_Tau0_pT_dists.pdf}
%\end{tabular}
%\vspace{0.3cm}
%\vspace{0.3cm}
%\begin{tabular}{cc}
%\includegraphics[width=.32\textheight, page=6]{plots/MiNNLO_Tau0_pT_dists.pdf}
%&
%\includegraphics[width=.32\textheight, page=4]{plots/MiNNLO_Tau0_pT_dists.pdf}
%\end{tabular}
\caption{\label{fig:jetdistributions} Comparison of \minnlo{}-$\Tau_0$ (green, dashed) and \minnlo{}-$\pt{}$ (blue, solid) predictions for differential distributions involving the hardest jet for Higgs (left) and Drell-Yan (right) production.}
\end{center}
\end{figure}
\clearpage}

We finish the validation of our \minnlo{}-$\Tau_0$ implementation by considering distributions that require the presence of at least one jet in the final state. 
Such distributions, by construction, are only NLO accurate, and accordingly \minnlo{} predictions have the same formal accuracy as \minlo{} ones.
However, since the kinematical origin of the NNLO corrections added through the \minnlo{} procedure corresponds to that of the Born phase space without any
extra jets, building a consistent NNLO+PS generator requires some form of spreading of these distributions in one-jet phase space. Indeed, such spreading 
is implemented in the \minnlo{} method for both the $\pt$ and the $\Tau_0$ matching, see $F^{\tmop{corr}}(\PhiBJ)$ in \eqn{eq:Bbar} and \eqn{eq:BbarTau0}, 
respectively.
It is obvious that such spreading for \minnlo{}-$\pt$ and \minnlo{}-$\Tau_0$ may have different effects in the one-jet phase space. Nevertheless, the predictions
are expected to be in reasonable agreement with each other and with the \minlo{} ones within the respective scale uncertainties. While such validation has
been performed (for various processes) for the \minnlo{}-$\pt$ implementation by comparing to \minlo{} results, here we need to perform such validation also
for our new \minnlo{}-$\Tau_0$ implementation.

To this end, we compare \minnlo{}-$\Tau_0$ predictions to \minnlo{}-$\pt$ ones for distributions in the one-jet phase space in \fig{fig:jetdistributions}
for both Higgs-boson production and Drell-Yan production. In particular, we show the distribution in the transverse momentum ($\ptjone$) and 
rapidity ($\yjone$) of the leading jet.
The results for Higgs and Drell-Yan production turn out to be rather different. While for Drell-Yan, by and large, \minnlo{}-$\Tau_0$ and \minnlo{}-$\pt$
are consistent with each other within uncertainties, with acceptable (and not unexpected) differences in terms of shape, for Higgs-boson 
production we observe much larger differences between the two predictions.
In particular, we observe that the \minnlo{}-$\Tau_0$ prediction is about $20\%$ larger at $\ptj  > 70$~GeV and at central rapidities of the leading jet.
Those differences are not fully covered by the scale uncertainty bands, which are at the $10\%$ level, and they 
are already present at the LHE level, i.e.\ before shower effects are included.
We found that modifying the spreading function of the inclusive NNLO correction for \minnlo{}-$\Tau_0$ in the one-jet phase space can have an impact
on those distributions in the case of Higgs-boson production. However, while variations of the spreading function partially mitigate the differences with \minnlo{}-$\pt$, 
they do not eliminate them completely.

We would like to stress that as far as NNLO accuracy is concerned, the \minnlo{}-$\Tau_0$ implementation provides correct results
also for Higgs-boson production. However, since our aim is a full fledged NNLO+PS Monte-Carlo generator, the issues observed in 
jet-related quantities of the \minnlo{}-$\Tau_0$ matching pose a certain level of concern for Higgs-boson production. 
Nevertheless, we reckon that the new results for Drell-Yan production are very encouraging and make a \minnlo{} implementation based on jettiness
worthwhile, also in view of moving towards higher jet multiplicities. Moreover, one should bear in mind that the present paper should be considered 
a first step towards NNLO+PS matching for processes with an extra jet in the final state. Eventually, the relevant ingredients for NNLO matching for new one-jet 
resolution variables will become available, which renders the present and forthcoming studies crucial.

%----------------------------------------------------------------------------
\subsection{Comparison to GENEVA results for Drell-Yan production}
\label{sec:MiNNLOvsGeneva}
We continue our study of phenomenological results by presenting a first direct comparison between the predictions 
from our \minnlo{} generators to the ones from the {\sc Geneva} generators \cite{Alioli:2013hqa,Alioli:2021qbf}. Also in the case of 
the {\sc Geneva} method there exist two different implementations for the Drell-Yan process, one using $\Tau_0$ as the matching 
variable (the default choice for {\sc Geneva} predictions so far) \cite{Alioli:2013hqa} and one using $\pt{}$ as the matching variable. The {\sc Geneva}-$\pt{}$ implementation employs the resummed $\pt{}$ spectrum
obtained through {\sc RadISH}~\cite{Monni:2016ktx,Bizon:2017rah} by means of an interpolation of a grid fully differential in the degrees of freedom of the Born phase space as well as in $\pt{}$. 
%\afterpage{
\begin{figure}[t]
\begin{center}\vspace{-0.2cm}
\begin{tabular}{cc}
\includegraphics[width=.32\textheight, page=1]{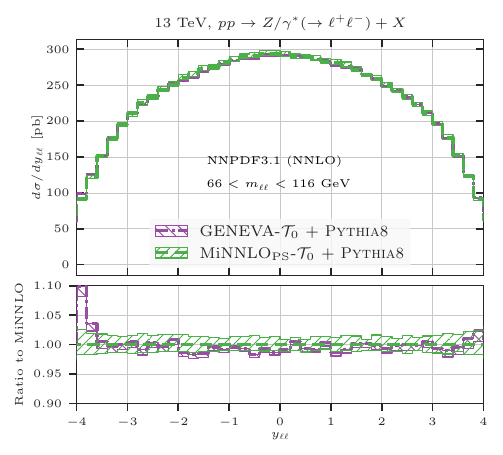}
&
\includegraphics[width=.32\textheight, page=2]{plots/GENEVA_MiNNLO_Tau0_dists.pdf}
\end{tabular}\vspace{0.3cm}
\begin{tabular}{cc}
\includegraphics[width=.32\textheight, page=5]{plots/GENEVA_MiNNLO_Tau0_dists.pdf}
&
\includegraphics[width=.32\textheight, page=4]{plots/GENEVA_MiNNLO_Tau0_dists.pdf}
\end{tabular}
%\vspace{0.3cm}
%\begin{tabular}{cc}
%\includegraphics[width=.32\textheight, page=6]{plots/GENEVA_MiNNLO_Tau0_ATLAS_dists.pdf}
%&
%\includegraphics[width=.32\textheight, page=4]{plots/GENEVA_MiNNLO_Tau0_ATLAS_dists.pdf}
%\end{tabular}
\caption{\label{fig:genevatau0} Comparison of \minnlo{}-$\Tau_0$ (green, dashed) and {\sc Geneva}-$\Tau_0$ (purple, dot-dashed) predictions for differential distributions in the phase-space of the leptons in Drell-Yan production.}
\end{center}
\end{figure}
%\clearpage}

%\afterpage{
\begin{figure}[t]
\begin{center}\vspace{-0.2cm}
\begin{tabular}{cc}
\includegraphics[width=.32\textheight, page=1]{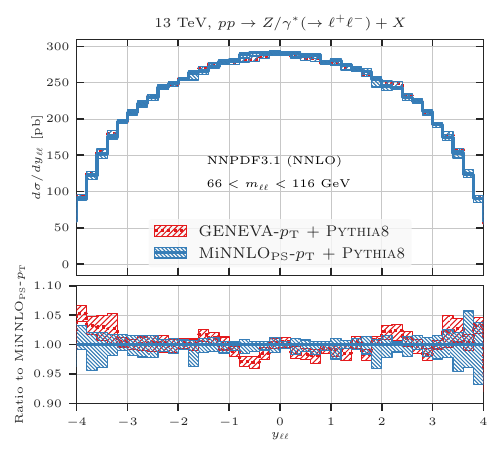}
&
\includegraphics[width=.32\textheight, page=2]{plots/GENEVA_MiNNLO_pT_dists.pdf}
\end{tabular}\vspace{0.3cm}
\begin{tabular}{cc}
\includegraphics[width=.32\textheight, page=5]{plots/GENEVA_MiNNLO_pT_dists.pdf}
&
\includegraphics[width=.32\textheight, page=3]{plots/GENEVA_MiNNLO_pT_dists.pdf}
\end{tabular}
%\vspace{0.3cm}
%\begin{tabular}{cc}
%\includegraphics[width=.32\textheight, page=6]{plots/GENEVA_MiNNLO_pT_ATLAS_dists.pdf}
%&
%\includegraphics[width=.32\textheight, page=4]{plots/GENEVA_MiNNLO_pT_ATLAS_dists.pdf}
%\end{tabular}
\caption{\label{fig:genevapt} Comparison of \minnlo{}-$\pt$ (blue, solid) and {\sc Geneva}-$\pt$ (red, dotted) predictions for differential distributions in the phase-space of the leptons in Drell-Yan production.}
\end{center}
\end{figure}
%\clearpage}

\Figs{fig:genevatau0} and~\ref{fig:genevapt} show the comparison of \minnlo{} and {\sc Geneva} predictions in the phase space of the two final-state leptons
using as matching variables $\Tau_0$ and $\pt$, respectively. Given that \minnlo{} and {\sc Geneva} treat terms beyond accuracy differently, we do not expect
a one-to-one correspondence between their results, but rather that their NNLO predictions agree within the associated scale uncertainties. Indeed, we observe 
for both the $\Tau_0$ and $\pt$ results that  \minnlo{} and {\sc Geneva} predictions are in full agreement within uncertainties.
Note that for Born-level observables the public version of {\sc Geneva}-$\mathcal T_0$ allows one to calculate the uncertainty only with a three-scale variation as described in~\citere{Alioli:2015toa}, which explains why the GENEVA-$\mathcal T_0$ uncertainties are somewhat smaller than the 
\minnlo{}-$\mathcal T_0$ ones. Similarly to before, the 
largest relative differences appear for $\ptlp>m_Z/2$, with a visible shape distortion around the threshold, which is sensitive to soft-gluon 
effects and therefore to the specific matching scheme. We recall that these differences as well as the widening of the uncertainty bands for $\ptlp>m_Z/2$
is expected, since the accuracy of the predictions is effectively reduced to NLO.

\subsection{Comparison against ATLAS and CMS data for Drell-Yan production}
\label{sec:data}

\begin{figure}[t]
\begin{center}\vspace{-0.2cm}
\begin{tabular}{cc}
\includegraphics[width=.32\textheight]{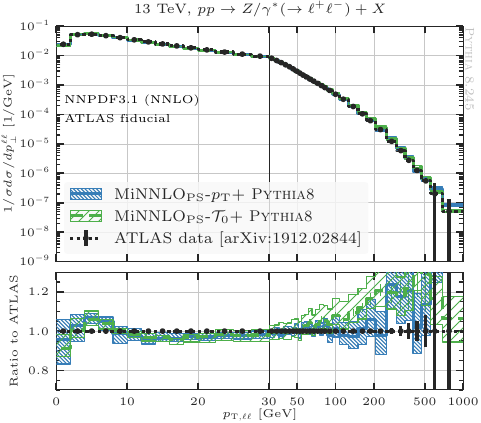}
&
\includegraphics[width=.32\textheight]{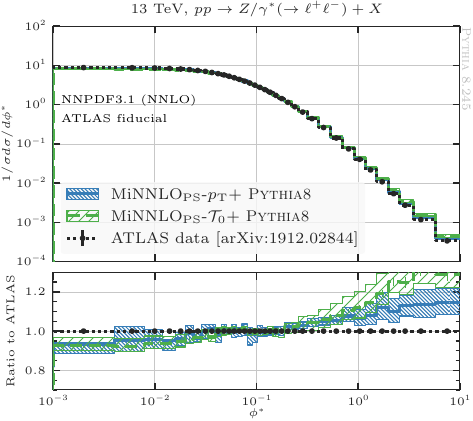}
\end{tabular}
%\vspace{0.3cm}
%\begin{tabular}{cc}
%\includegraphics[width=.32\textheight]{plots/MiNNLO_Tau0_pT_CMS_ptZ.pdf}
%&
%\includegraphics[width=.32\textheight]{plots/MiNNLO_Tau0_pT_CMS_phistar.pdf}
%\end{tabular}
%\includegraphics[width=.32\textheight]{plots/MiNNLO_Tau0_pT_CMS_YV.pdf}
\caption{\label{fig:atlas} Comparison of \minnlo-$\pt$ (blue, solid) and \minnlo-$\Tau_0$ (green, dashed) against ATLAS data from \citere{ATLAS:2019zci} for the transverse momentum of the lepton pair (left) and the Collins-Soper angle $\phi^*$ (right), as defined in the main text.}
\end{center}
\end{figure}
\begin{figure}[t]
\begin{center}\vspace{-0.2cm}
%\begin{tabular}{cc}
%\includegraphics[width=.32\textheight]{plots/MiNNLO_Tau0_pT_ATLAS_ptZ.pdf}
%&
%\includegraphics[width=.32\textheight]{plots/MiNNLO_Tau0_pT_ATLAS_phistar.pdf}
%\end{tabular}\vspace{0.3cm}
\begin{tabular}{cc}
\includegraphics[width=.32\textheight]{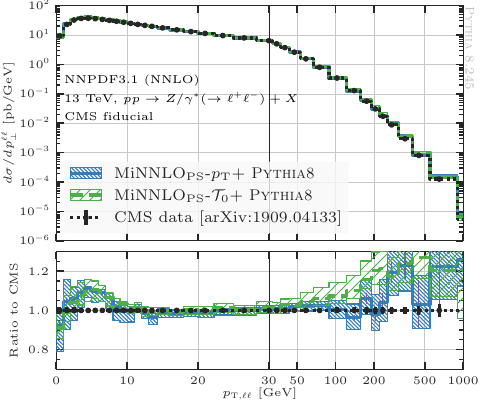}
&
\includegraphics[width=.32\textheight]{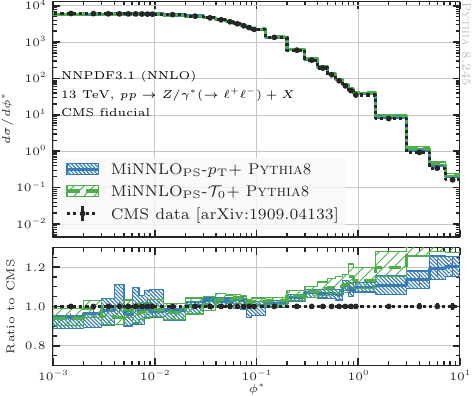}
\end{tabular}
\includegraphics[width=.32\textheight]{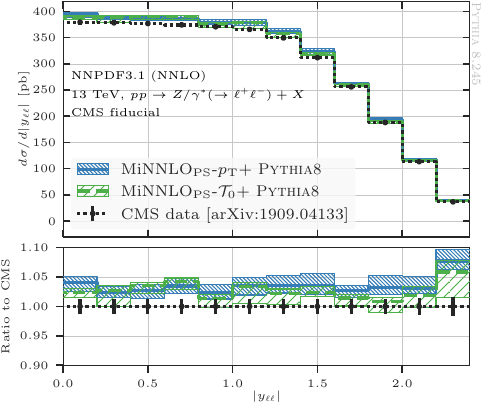}
\caption{\label{fig:cms}Comparison of \minnlo-$\pt$ (blue, solid) and \minnlo-$\Tau_0$ (green, dashed) against CMS data from ref.~\cite{CMS:2019raw} for the transverse momentum of the lepton pair (upper left) and the Collins-Soper angle $\phi^*$ (upper right), as defined in the main text, and the $Z$ boson rapidity distribution (bottom plot).}
\end{center}
\end{figure}

In this section we compare our \minnlo{} predictions for Drell-Yan production against ATLAS and CMS data.
  We consider the recent ATLAS analysis presented in \citere{ATLAS:2019zci}, where results for the transverse momentum of the dilepton
  system ($\ptll$) and a variation of the Collins-Soper angle ($\phi^*$) are shown. The angle $\phi^*$ is defined as
\begin{equation}
  \phi^* = \tan\left(\frac{\pi-\Delta\phi}{2}\right)\sin\left(\theta^*\right)\,,\quad
  \cos\left(\theta^*\right)=\tanh\left(\frac{\Delta \eta}{2}\right)\,,
\end{equation}  
where $\Delta \eta$ and $\Delta \phi$ are the differences in
pseudorapidity and azimuthal angle between the two leptons.
As for the CMS data, we consider the analysis presented in \citere{CMS:2019raw} where, besides results for $\ptll$ and $\phi^*$,
also the rapidity distribution of the dilepton system ($\yll$) is shown. The two analyses use similar fiducial cuts, which
are reported in table \ref{tab:cuts}.
\renewcommand{\baselinestretch}{1.5}
\begin{table}[]
\centering
 \begin{tabular}{c|c}
    ATLAS~\cite{ATLAS:2019zci} & CMS~\cite{CMS:2019raw} \\
   \hline
   $\ptl>27$~GeV &  $\ptl>25$~GeV \\
   $|\etal|< 2.5$ & $|\etal|< 2.4$ \\
   $66~\mathrm{GeV}<\mll<116~\mathrm{GeV}$ & $|\mll-m_Z|\,\mathrm{GeV}<15 \,\mathrm{GeV}$
 \end{tabular}
 \renewcommand{\baselinestretch}{1.0}
 \caption{Fiducial cuts used in the ATLAS and CMS analyses.}
 \label{tab:cuts}
\end{table}
\renewcommand{\baselinestretch}{1.0}
In \fig{fig:atlas} we present a comparison between \minnlo-$\pt$ (blue,
solid) and \minnlo-$\Tau_0$ (green, dashed) predictions with ATLAS data. 
As for the transverse momentum of the dilepton system, our \minnlo{} predictions
are in good agreement with data throughout the entire spectrum.
For very small values of $\ptll$ ($\ptll<10$~GeV), we observe
a slight difference in shape between the \minnlo{} curves and data, which is however not unexpected as this region is sensitive to soft-collinear radiation and requires an accurate resummation of large logarithmic terms.
At large $\ptll$ values, both generators are
NLO accurate only, which is reflected in the enlarged theory uncertainty
bands. In this high-$\ptll$ region theory predictions
tend to overestimate data but the agreement remains good, at 1--2~$\sigma$ level.
As for the angle $\phi^*$, we observe that both
generators agree rather well with data (1--2~$\sigma$ level), but data tend to
fall more sharply at large $\phi^*$ values. In \fig{fig:cms} we present a comparison with CMS data.
In this comparison we observe
the same relative behaviour as with ATLAS data for both $\ptll$ and $\phi^*$, so the same conclusions hold.
Moreover, we present results for the rapidity distribution of the reconstructed $Z$ boson $\yll$, for which we observe an excellent
description of the data with both \minnlo{}-$\pt$ and \minnlo{}-$\Tau_0$, with a discrepancy of a few percent only, relatively flat across the whole rapidity range.

%============================================================================
\section{Conclusions}
\label{sec:conclusions}
%\label{sec:higgsplusjet}

We have presented the derivation of the \minnlo{} formalism using jettiness as resummation variable. 
This calculation opens the door to consider NNLO+PS matching for processes with additional jets in the final state. As a proof-of-concept
we have performed a complete implementation of the \minnlo{} approach using the $0$-jettiness variable for 
colour-singlet production. Specifically, we have considered Higgs-boson production and Drell-Yan production as a first implementation.

We have validated our \minnlo{}-$\Tau_0$ implementation against the existing \minnlo{}-$\pt$ generators, finding excellent agreement among their
predictions within uncertainties for NNLO-accurate observables, including the total inclusive cross section and differential distributions, both 
for Higgs-boson and Drell-Yan production.
On the other hand, for jet-related quantities in Higgs-boson production we observe larger differences between  \minnlo{}-$\Tau_0$ and \minnlo{}-$\pt$ predictions. Those 
are present both at Les-Houches-Event level and after the inclusion of parton shower effects, but their relative difference remains almost identical before and after showering.
The corresponding jet-related results for the Drell-Yan process are in good agreement
between the \minnlo{}-$\Tau_0$ and \minnlo{}-$\pt$ implementations.

We then continued by presenting the first direct comparison of \minnlo{} predictions with existing results from the {\sc Geneva} generators 
for Drell-Yan production. For this process both \minnlo{} and {\sc Geneva} implementations exist using $\Tau_0$ and $\pt$ as a matching 
variable. We have shown that for NNLO-accurate quantities \minnlo{} and {\sc Geneva} results are in excellent agreement within the 
respective higher-order uncertainties, which is not unexpected given that both approaches yield NNLO accurate predictions.

Finally, we have compared  \minnlo{}-$\Tau_0$ and \minnlo{}-$\pt$ predictions to Drell-Yan data recorded by the ATLAS and by the CMS
collaboration. By and large, we have found \minnlo{} predictions to be in remarkable agreement with the experimental data, with differences mostly within 
one standard deviation.

This work is a first important step towards NNLO+PS matching for processes with massless partons in the final state, such as Higgs plus 
jet and vector boson plus jet production, which has not been achieved for any such process to date. The implementation of the \minnlo{} approach
based on $\Tau_0$ has been fully worked out, implemented and validated.
We have also fully worked out all necessary formulae to obtain NNLO+PS accurate predictions for colour singlet plus jet production using $\Tau_1$ as a resummation variable. All equations are reported in Appendix~\ref{sec:higgsplusjet}.
The complete implementation for $\Tau_1$ and application to a corresponding process, such as Higgs plus jet or Drell-Yan plus jet production, is left to future work.

%  We anticipate that the differences on the distributions we consider are rather moderate, thereby partially justifying the choice of postponing this issue to future work.

%============================================================================
\section*{Acknowledgements}

We would like to thank Simone Alioli, Pier Francesco Monni and Paolo Nason for fruitful discussions, and we are grateful to Simone Alioli for comments on the manuscript.
We also thank Kenneth Long for providing us details on the \textsc{Pythia} tune used for the \minnlo{} predictions by CMS and for useful discussion on the experimental analysis.
L.R. has been supported by the SNSF under contract PZ00P2 201878.
S.Z. has been supported by the International Max Planck Research School (IMPRS) on ``Elementary Particle Physics''. The research of S.Z. has also been supported by the European Research Council
(ERC) under the European Union's Horizon 2020 research and innovation program (grant agreement No. 788223, PanScales,
and grant agreement No. 804394, HipQCD) and by the Science and Technology Facilities Council (STFC) under grant ST/T000864/1. 
\appendix

%============================================================================
\section{Explicit formulae for \minnlo{}-$\Tau_0$}
\label{app:constants}

  In this appendix,
  we collect explicit formulae of relevant perturbative ingredients we used or obtained in the main sections of this paper.

\paragraph{Notation.} The QCD $\beta$ function is expanded in terms of the strong coupling $\as$ as
\begin{align} \label{eq:conventions_2}
 \beta[\as(\mu)] &
 = \frac{\df\as(\mu)}{\df\ln\mu^2}
 = - \as(\mu) \sum_{n=0}^\infty \beta_n \, [\as(\mu)]^{n+1}
\,,\end{align}
where the coefficients $\beta_n$ are given by~\cite{Tarasov:1980au, Larin:1993tp}
\begin{align} \label{eq:beta_coeffs}
 \beta_0 &
 = \frac{11 C_A - 2 n_f}{12 \pi}
\,,\nn\\
 \beta_1 &
 = \frac{17 C_A^2 - (5 C_A + 3 C_F) n_f}{24 \pi^2}
\,,\nn\\
 \beta_2 &
 = \frac{2857 C_A^3 + (54 C_F^2 - 615 C_A C_F -1415 C_A^2) n_f + (66 C_F + 79 C_A) n_f^2}{3456 \pi ^3}
\,.\end{align}

The DGLAP evolution kernels are defined by
\begin{align}
 \frac{\partial f_i(x, \mu)}{\partial \ln\mu^2} &
 = \sum_j \int_x^1 \frac{\df x'}{x'} \hat P_{ij}[x', \as(\mu)] f_j\left(\frac{x}{x'}, \mu\right)
\,,
\end{align}
with the expansion of the splitting function given in \eqn{eq:expP}. We refer to \citere{Monni:2019whf} for explicit expressions for the
coefficients $\hat P^{(n)}(x)$.

  Given a generic function $F$ depending on the renormalization scale $\mu$ through the running coupling $\as$,
  we define its perturbative expansion in terms of $\as / (2\pi)$ as:
\begin{align} \label{eq:conventions_1}
 F(\dots, \mu) = \sum_{n=0}^\infty F^{(n)}(\dots) \left[\frac{\as(\mu)}{2\pi}\right]^n
\,.\end{align}
We note that this convention differs from the standard one used in the SCET literature.
For a collection of all required ingredients in the standard SCET notation, see e.g.~\citere{Billis:2019vxg}.

\paragraph{$\Tau_0$ Sudakov form factor.} The \minnlo{}-$\Tau_0$ Sudakov form factor presented in \eqref{eq:Tau0_L_S} depends on the A and B coefficients
  defined as in \eqref{eq:def_B_Tau0}
\begin{align} \label{eq:def_B_Tau0_app}
 A(\as) = \GammaC(\as)
\,,\quad
 B_{F}(\as) = \frac12 \gamma_F(\as) - \beta(\as) \frac{\df\ln F(\as)}{\df\as}
 \,,\end{align}
where $F$ can either be related to the hard or the soft functions.
These coefficients admit the following perturbative expansion:
\begin{align}
 A(\as) = \sum_{n=1}^\infty A^{(n)} \left(\frac{\as}{2\pi}\right)^n
\,,\qquad
 B_F(\as) = \sum_{n=1}^\infty B_F^{(n)} \left(\frac{\as}{2\pi}\right)^n
\,.\end{align}
The $A$ coefficient is identical to the cusp anomalous dimension.
Up to three loops, its coefficients obey Casimir scaling, i.e. $\GammaC^{(n)}\propto C$, where $C$ is the Casimir of the given representation, for any given perturbative order $n$. 
{Up to the third order, the coefficients $A^{(n)}$ are given by~\cite{Korchemsky:1987wg, Moch:2004pa, Vogt:2004mw}
\begin{align} \label{eq:A_coeffs}
 A^{(1)} &
 \equiv \Gamma_C^{(1)}
 = 2 C
\,,\nn\\
 A^{(2)} &
 \equiv \Gamma_C^{(2)}
 = C \biggl[ C_A \Bigl(\frac{67}{9} - 2 \zeta_2 \Bigr) - \frac{10}{9} n_f \biggr]
\,,\nn\\
 A^{(3)} &
 \equiv \Gamma_C^{(3)}
 = C \biggl[ C_A^2 \Bigl(\frac{245}{12} - \frac{268}{18}\zeta_2 + \frac{22}{6}\zeta_3 + 11 \zeta_4 \Bigr)
   + C_A n_f \Bigl(-\frac{209}{54} + \frac{20}{9}\zeta_2 - \frac{14}{3}\zeta_3 \Bigr)
   \nn \\ & \hspace{2.5cm}
   + C_F n_f \Bigl(-\frac{55}{12} + 4\zeta_3 \Bigr)
   - \frac{2}{27} n_f^2 \biggr]
\,,\end{align}
where $C = C_F$ for quark-induced processes and $C = C_A$ for gluon-induced processes.
Note that $A^{(1)}$ and $A^{(2)}$ agree with the corresponding expressions for $\pt{}$ resummation
given in \citere{Monni:2019whf},
while the expression of $A^{(3)}$ differs from the one for $\pt{}$ resummation, as given in \citere{Monni:2019whf}.

The $B$ coefficient for the hard function reads
\begin{align} \label{eq:def_B_H}
 B_H(\as) = 2 \gamma_H(\as) - \beta(\as) \frac{\df \ln H_{ab}(\as)}{\df \as}
 \,,\end{align}
where $H_{ab}$ is the hard function and $\gamma_H$ is the hard anomalous dimension, with $\gamma_H = \gamma^q_H (\gamma^g_H)$ for quarks (gluons).  
To the required order, they read~\cite{Fleming:2003gt,Becher:2006mr,Becher:2009th,Stewart:2010qs,Berger:2010xi}
\begin{align}
 2 \gamma_H^{q(1)} &
 = - 3 C_F
\,,\\\nn
 2 \gamma_H^{q(2)} &
 = C_F \left[
   C_F \left(6 \zeta_2 - 12 \zeta_3 - \frac{3}{4}\right)
   + C_A \left(-\frac{11}{2} \zeta_2 + 13 \zeta_3 - \frac{961}{108}\right)
   + n_f \left( \zeta_2  + \frac{65}{54}\right) \right]
\,,\\
 2 \gamma_H^{g(1)} &
 = - 4 \pi \beta_0
\,,\\\nn
 2 \gamma_H^{g(2)} &
= C_A^2 \left(\frac{11}{6 }\zeta_2 + \zeta_3 - \frac{346}{27}\right)
   + C_A n_f \left(\frac{64}{27} - \frac{\zeta_2}{3}\right) + C_F n_f
\,.\end{align}
The fixed-order expansion of \eq{def_B_H} thus reads
\begin{align} \label{eq:B_H_coeffs}
 B_H^{(1)} &
 = 2 \gamma_H^{(1)}
\,,\nn\\
 B_H^{(2)} &
 = 2 \gamma_H^{(2)}
   + 2 \pi \beta_0 H^{(1)}
\,,\end{align}
where $H^{(1)}$ refers to the one-loop coefficient of the corresponding hard function,
which is given below for Higgs-boson and Drell-Yan production.

For the $\Tau_0$ soft function, the $B$ coefficient is defined as
\begin{align} \label{eq:def_B_S}
 B_S(\as) = \frac12 \gamma_S(\as) - \beta(\as) \frac{\df \ln S(\as)}{\df \as}
\,,\end{align}
where $\gamma_S$ is the anomalous dimension of the 
soft function $S$. Up to second order, its expansion coefficients obey Casimir scaling and read
\begin{align}
  \gamma_S^{(1)} & = 0  \,,\\\nn
  \gamma_S^{(2)} & = C \left[ C_A \left(\frac{11}{3}\zeta_2 + 14 \zeta_3 - \frac{404}{27}\right)+ n_f \left(\frac{56}{27} - \frac{2}{3}\zeta_2 \right) \right] \,,
\end{align}
 The fixed-order expansion of \eq{def_B_S} reads
\begin{align}
 B_S^{(1)} &= 0
%%%
\,,\nn\\
%%%
 B_S^{(2)} &
 = C \left[ C_A\left(7 \zeta_3 - \frac{16}{9} \right)
              - 2 \pi \beta_0 \left(2\zeta_2 + \frac{28}{9} \right) \right]
\,.\end{align}
Here, we used the one-loop result for $S$ given in \eq{soft_coeffs}.

We now present explicit expressions for the hard, beam and soft functions.
\paragraph{$\Tau_0$ hard function.}
The hard function is defined in \eq{H_S_canonical} as
\begin{align}
 H_{ab}(Q, \mu=Q) = 1 + \frac{\as(Q)}{2\pi} H_{ab}^{(1)} + \left[\frac{\as(Q)}{2\pi}\right]^2 H_{ab}^{(2)}  + \cO(\as^3)
\,.\end{align}
The coefficients $H^{(1)}$ and $H^{(2)}$ for Higgs and Drell-Yan production are identical to those
used in \citere{Monni:2019whf} without including the shift $\Delta H^{(2)}$ from momentum-space resummation.
For completeness, we repeat them here.

The hard function for Higgs production in the $m_t\to\infty$ limit is obtained by combining the IR-finite gluon form factor,
which is known up to three loops~\cite{Harlander:2000mg, Gehrmann:2005pd, Moch:2005tm, Baikov:2009bg, Lee:2010cga, Gehrmann:2010ue},
with the Wilson coefficient from integrating out the top quark,
which itself is known up to four loops~\cite{Chetyrkin:1997un,Schroder:2005hy,Chetyrkin:2005ia}.
Here, we only need the results up to two loops, which are given by
\begin{align} \label{eq:hard_coeffs_Higgs}
 H_{gg}^{(1)} &
 = C_A \left(5 + \frac{7}{6} \pi^2 \right) - 3 C_F
\,,\nn\\
 H_{gg}^{(2)} &
 = \left(\frac{7}{2}C_A^2 - \frac{11}{2} C_A C_F +2 C_F n_f\right) \log\frac{m_H^2}{m_t^2}
 +C_A^2 \left(\frac{755 \pi^2}{72} - \frac{143 \zeta_3}{18} + \frac{37 \pi^4}{72} + \frac{23827}{648}\right)
 \nn\\&\quad
 + C_A \left[ C_F \left(-\frac{145}{6} - \frac{7}{2} \pi^2 \right)
 + n_f \left(-\frac{23}{9}\zeta_3 - \frac{25}{36}\pi^2 - \frac{2255}{324}\right) -\frac{5}{24} \right]
 \nn\\&\quad
 + C_F \left[ 9 C_F + n_f \left(4 \zeta_3-\frac{41}{6}\right) - \frac13 \right]
\,.\end{align}
For Drell-Yan, the hard function is obtained from the IR-finite quark vector form factor
which is known up to three loops~\cite{Kramer:1986sg, Matsuura:1987wt, Matsuura:1988sm, Gehrmann:2005pd, Moch:2005tm, Moch:2005id, Baikov:2009bg, Lee:2010cga, Gehrmann:2010ue}.
Starting at $\cO(\as^2)$ there are also nonvanishing singlet contributions from the axial anomaly,
which are known to have a small effect on the cross section~\cite{Dicus:1985wx, Hamberg:1990np}
and are thus neglected here. (For a detailed discussion of its inclusion in the hard function, see e.g.~\citere{Ju:2021lah}.)
The hard function coefficients are
\begin{align} \label{eq:hard_coeffs_DrellYan}
 H_{q\bar q}^{(1)} &
 = C_F \left( \frac{7}{6} \pi^2 - 8 \right)
\,,\nn\\
 H_{q\bar q}^{(2)} &
 =  C_F^2 \left( - \frac{83}{12} \pi^2  - 15 \zeta_3 + \frac{67}{120} \pi^4 + \frac{511}{16}\right)
 + C_A C_F \left(\frac{1061}{216}\pi^2 + \frac{313}{18} \zeta_3 - \frac{2}{45}\pi^4 - \frac{51157}{1296}\right)
 \nn\\&\quad
 + C_F n_f \left( - \frac{91}{108}\pi^2 + \frac{\zeta_3}{9} + \frac{4085}{648}\right)
\,.\end{align}
Setting $N_c = 3$ and $n_f = 5$, \eqs{hard_coeffs_Higgs}{hard_coeffs_DrellYan}
reproduce eqs.~(B.10) and (B.12) in \citere{Monni:2019whf}.

\paragraph{$\Tau_0$ beam functions.}
The beam functions for leptonic $\Tau_0$ are defined in \eqs{beam_matching}{beam_constants} as
\begin{align}
 B_i\Bigl(\frac{y}{Q}, x, \mu_B\Bigr) &
 = \sum_j \int_x^1 \frac{\df x'}{x'} \bar C_{ij}\Bigl(\frac{y}{Q}, x', \mu_B\Bigr) f_j\Bigl(\frac{x}{x'}, \mu_B\Bigr)
\,,\nn\\
 \bar C_{ij}(x, \mu_B)
 &\equiv \bar C_{ij}\Bigl(\frac{y}{Q}, x, \mu_B \Bigr)
 = \delta_{ij} \delta(1-z) + \sum_{n=1}^\infty \left[\frac{\as(\mu_B)}{2\pi}\right]^n \bar C_{ij}^{(n)}(z)
\,,\end{align}
where $\mu_B = \sqrt{Q y_0 / y}$ is fixed to minimize all logarithms in Fourier space.
At two loops, the beam functions have been calculated in \citeres{Stewart:2010qs, Gaunt:2014xga, Gaunt:2014cfa}
in momentum space. To obtain the functions $\bar C_{ij}$, one has to first perform
the Fourier transform as defined in \eq{FT}, and then choose $\mu_B = \sqrt{Q y_0 / y}$
to eliminate explicit logarithms. Alternatively, one can obtain them directly
from the results provided in \citere{Ebert:2020unb} in Fourier space upon setting the
logarithm $L_y = 0$ in there.%
\footnote{Note that the results in \citere{Ebert:2020unb} are expanded in $\as/\pi$ instead of $\as/(2\pi)$.}
It is then trivial to combine the results for $\bar C_{ij}$ with the soft function given below
to obtain the coefficients $C_{ij}$ as defined in \eq{beam_matching_2}.
Since the $\bar C_{ij}$ are rather lengthy and provided as Mathematica files with \citere{Ebert:2020unb},
we do not provide explicit expressions here.

\paragraph{$\Tau_0$ soft function.}
The soft function is defined in \eq{H_S_canonical} as
\begin{align}
 S(y_0/y) = 1 + \frac{\as(y_0/y)}{2\pi} S^{(1)} + \left[\frac{\as(y_0/y)}{2\pi}\right]^2 S^{(2)}  + \cO(\as^3)
\,.\end{align}
The two loop results have been calculated in momentum space in~\citeres{Kelley:2011ng, Monni:2011gb, Hornig:2011iu, Kang:2015moa}.
Taking cross terms induced by the Fourier transform into account, we obtain
\begin{align} \label{eq:soft_coeffs}
 S^{(1)} &
 = - \frac{\pi^2}{2} C
\,,\\\nn
 S^{(2)} &
 = C \left[ C_A \left( - \frac{\pi^2}{9} + \frac{7}{30} \pi^4 -\frac{160}{27} \right)
   + 4 \pi \beta_0 \left( - \frac{77}{72} \pi^2 + \frac{13}{6} \zeta_3 - \frac{5}{27}\right) \right]
   + \frac{\pi^4}{8} C^2
\,,\end{align}
where $C = C_F$ for quark annihilation and $C = C_A$ for gluon fusion.

\section{Phase space parametrisation for the $D^{(\geq 3)}[\mathcal T_0]$ term}\label{app:spreading}

In this appendix we discuss the parametrisation of the factor $F^{\tmop{corr}}(\PhiBJ)$ in \eqn{eq:BbarTau0}.
The starting point is the \POWHEG{} projection of the FJ phase space $\Phi_{\rm FJ}$ onto the F phase space $\Phi_{\rm F}$ for initial state radiation.
The $\Phi_{\rm FJ}$ phase space can be expressed in the following factorised form (see 5.1.1~of \cite{Frixione:2007vw} for additional details)
\begin{equation}
	\df \Phi_{\rm FJ} = \df \Phi_{\rm F}  \df\Phi_{\rm rad}, \qquad \df\Phi_{\rm rad} = \frac{s}{(4\pi)^3} \frac{\xi}{1-\xi} \df\xi\df\phi\df y
\end{equation}
where $s$ is the square of the total incoming energy and we have introduced the variables
\begin{align} \label{eq:vars}
 \xi = \frac{2 k^0}{\sqrt s}
\,,\qquad
 y = \cos\theta
\,,\qquad
 k_T^2 = \frac{s}{4} \xi^2 (1-y^2)
\,,\end{align}
where $k^0$, $\theta$ and $\phi$ are the energy, the scattering angle and the azimuth of the radiated parton, respectively, defined in the centre-of-mass frame of the FJ system.

For a single real emission the definition of $\Tau_0$ in \eq{def_Tau0} yields
\begin{align} \label{eq:Tau0_vals_FKS_1}
 \Tau_0 = k_T e^{-|Y - \eta_k|}
\,,\end{align}
where $\eta_k$ is the rapidity of the emission in the laboratory frame.
Using the relation
\begin{align} \label{eq:eta_k}
 \eta_k = \frac12 \ln\frac{x_1 (1+y)}{x_2 (1-y)}
\,.\end{align}
we obtain
\begin{align} \label{eq:Tau0_vals_FKS_2}
 \Tau_0
 = k_T e^{-|\Delta Y(y) |}
\,,\qquad
 \Delta Y(y)
 = \frac12 \ln\frac{x_1 (1+y)}{x_2 (1-y)}
 - Y\,.\end{align}
In \eq{Tau0_vals_FKS_2}, it is important to note that $k_T$ and $x_{1,2}$ themselves
depend on $\xi$ and $y$.
Making all of this dependence manifest, we obtain
\begin{align} \label{eq:Tau0_vals_FKS_3}
 \Tau_0 &
 = \frac{M \xi }{2} \frac{\sqrt{1-y^2}}{\sqrt{1-\xi}}  e^{-|\Delta Y(\bar x_{1,2}, \xi, y) |}
\,,\nn\\
 \Delta Y(\bar x_{1,2}, \xi, y) &
 = \frac12 \ln\frac{1+y}{1-y}\frac{\bar x_1}{\bar x_2} \frac{2-\xi(1-y)}{2-\xi(1+y)}
 - Y
\,,\end{align}
where the barred momentum fractions are~\cite{Frixione:2007vw}
\begin{align} \label{eq:bar_x12}
 \bar x_1 = x_1 \sqrt{1-\xi} \sqrt{\frac{2-\xi(1+y)}{2-\xi(1-y)}}
\,,\qquad
 \bar x_2 = x_2 \sqrt{1-\xi} \sqrt{\frac{2-\xi(1-y)}{2-\xi(1+y)}}
\,.\end{align}

\subsection{Evaluation of $F_\ell^{\rm corr}$}

The factor $F^{\tmop{corr}}(\PhiBJ)$ is defined as
\begin{align} \label{eq:Fcorr_0}
 F_\ell^{\rm corr}(\Phi_{\rm FJ}) &
 = \frac{J_\ell(\Phi_{\rm FJ})}{\sum_{\ell'} \int\df\Phi'_\rad J_{\ell '}(\bar\Phi'_{\rm FJ}) \delta(\Tau_0 - \Tau_0') }
\,,\end{align}
where $\bar\Phi'_{\rm FJ} \equiv \Phi'_{\rm FJ} |_{\bar \Phi^\prime_{\rm F } = \Phi_{\rm F}} $. 

Using \eq{Tau0_vals_FKS_2}, we then obtain
\begin{align} \label{eq:Fcorr_1}
 \frac{J_\ell(\Phi_{\rm FJ})}{F_\ell^{\rm corr}(\Phi_{\rm FJ})} &
 = \sum_{\ell'} \int\df\xi\df\phi\df y \frac{s}{(4\pi)^3} \frac{\xi}{1-\xi} J_{\ell '}(\bar\Phi'_{\rm FJ})
   \delta\left(\Tau_0 - \frac{\xi \sqrt{s}}{2} \frac{\sqrt{1-y^2}}{e^{|\Delta Y(y)|}}\right)
\,,\end{align}
which is the straightforward extension of Eq.~(A.6) in \cite{Monni:2019whf} for $\Tau_0$.
After re-arranging the $\delta$ function and using  \eq{Tau0_vals_FKS_3} we have
\begin{align} \label{eq:Fcorr_4}
 \frac{J_\ell(\Phi_{\rm FJ})}{F_\ell^{\rm corr}(\Phi_{\rm FJ})} &
 = \frac{\Tau_0}{(2\pi)^2}  \sum_{\ell'} \int\df\xi\df y  J_{\ell '}(\bar\Phi'_{\rm FJ})
   \frac{\xi}{(1-\xi)^2} \delta\left(\frac{4\Tau_0^2}{M^2} - \frac{\xi^2}{1-\xi} \frac{1-y^2}{e^{2|\Delta Y(\bar x_{1,2}, \xi, y) |} } \right)
\,.\end{align}

In order to solve the $\delta$, we change the integration variables
from $(\xi, y)$ to $(k_T, \eta_k)$ using the relations
\begin{align} \label{eq:kt_eta}
 k_T = \frac{M}{2} \xi \frac{\sqrt{1-y^2}}{\sqrt{1-\xi}}
\,,\qquad
 \eta_k = \frac12 \ln\left[ \frac{\bar x_1}{\bar x_2} \frac{1+y}{1-y} {\frac{2-\xi(1-y)}{2-\xi(1+y)}} \right]
\,.\end{align}
The Jacobian of this variable transformation is given by
\begin{align}
  \df \xi \df y = \frac{2(1-\xi)^2}{M^2 \xi} \df k_T^2 \df \eta_k
\,.\end{align}
In terms of these variables, the FKS radiation phase space reads
\begin{align} \label{eq:phi_rad_2}
 \df\Phi_\rad &
 = \frac{s}{(4\pi)^3} \frac{\xi}{1-\xi} \df\xi\df\phi\df y
 = \frac{2}{(4\pi)^3} \df k_T^2 \df \eta_k \df \phi
\,.\end{align}
Applying this together with \eq{Tau0_vals_FKS_1} to \eq{Fcorr_0}, we obtain
\begin{align} \label{eq:Fcorr_5}
 \frac{J_\ell(\Phi_{\rm FJ})}{F_\ell^{\rm corr}(\Phi_{\rm FJ})} &
 = \sum_{\ell'} \int\df\Phi'_\rad J_{\ell '}(\bar\Phi'_{\rm FJ}) \delta\bigl(\Tau_0 - k_T e^{-|Y - \eta_k|}\bigr)
 \nn\\&
 = \sum_{\ell'} \int \frac{\df k_T^2 \df \eta_k}{(4\pi)^2} J_{\ell '}(\bar\Phi'_{\rm FJ}) \delta\bigl(\Tau_0 - k_T e^{-|Y - \eta_k|}\bigr)
 \nn\\&
 = \sum_{\ell'} \frac{\Tau_0}{8 \pi^2} \int\!\df \eta_k \, J_{\ell '}(\bar\Phi'_{\rm FJ}) e^{2|Y - \eta_k|}
\,,\end{align}

In order to evaluate $J_{\ell'}$ we now need
  to find the kinematic bounds of $\eta_k$, and express $y$ and $\xi$
    as a function of $k_T$ and $\eta_k$.
To see that the inverse of \eq{kt_eta} is unique, consider that
\begin{align}
 \frac{\df \eta_k}{\df y} &
 = \frac{4 - 2 \xi (1+y^2)}{4 (1 - y^2) (1 - \xi) + (1 - y^2)^2 \xi^2}
 > 0
\,,\end{align}
where the inequality follows since $-1\le y \le 1$ and $0 \le \xi \le 1$. It is useful to introduce the parameters
\begin{align}
 \kappa = \frac{\bar x_2}{\bar x_1} e^{2 \eta_k} = e^{2 (\eta_k - Y)}
\,,\qquad
 \beta = 2 + \frac{1+\kappa}{\sqrt\kappa} \sqrt{1 + \frac{M^2}{k_T^2}}
\,,\end{align}
through which we can express $\xi$ and $y$ as
\begin{align} \label{eq:xi_eta}
 \xi = 1 - \frac{1}{1 + \beta \frac{k_T^2}{M^2}}
\,,\qquad
 y = \frac{\kappa - 1}{\kappa+1} \left(1 - \frac{2}{\beta}\right)
\,.\end{align}
Note that  $\kappa \ge 0$ and $\beta \ge 2$,
and hence \eq{xi_eta} always obeys the obvious bounds $0<\xi<1$ and $-1<y<1$.
In order to identify the kinematic bounds on $k_T, \eta_k$
we insert \eq{xi_eta} into \eq{bar_x12},
\begin{align} \label{eq:x12_FKS}
 x_{1,2} &
 = \bar x_{1,2} \left( \sqrt{1+\eps^2} + \eps \kappa^{\pm 1/2}\right)
 \nn\\&
 = \bar x_{1,2} \left( \sqrt{1+\eps^2} + \eps \, e^{\pm (\eta_k - Y)} \right)
\,,\end{align}
where $\eps = k_T/M$. \Eq{x12_FKS} clearly implies $x_{1,2} \ge \bar x_{1,2}$,
so we only need to solve for the constraints $x_{1,2} \le 1$.
Since $x_{1,2}$ are monotonically growing with $\eps$, we easily find
\begin{align} \label{eq:eps_max}
 \eps = \frac{k_T}{M} \le \eps_{\rm max}
\,,\quad
 \eps_{\rm max}
 = \min\left\{ \frac{\sqrt{1 - (1-\kappa) \bar x_1^2} - \sqrt{\kappa}}{\bar x_1 (1-\kappa)}
   \,,\, (\bar x_1 \to \bar x_2, \kappa \to \kappa^{-1}) \right\}
   \,.\end{align}
Inserting this constraint into \eq{Fcorr_5}, we obtain
\begin{align} \label{eq:Fcorr_6}
 \frac{J_\ell(\Phi_{\rm FJ})}{F_\ell^{\rm corr}(\Phi_{\rm FJ})} &
 = \sum_{\ell'} \frac{\Tau_0}{8 \pi^2} \int\!\df \eta_k \, J_{\ell '}(\bar\Phi'_{\rm FJ}) e^{2|Y - \eta_k|}
   \Theta\left( \frac{M \eps_{\rm max}}{\Tau_0 e^{|Y - \eta_k|}} - 1 \right)
\,.\end{align}

\subsection{Analytic solution of the phase-space constraints}

We can finally derive an analytic solution to this phase space bound.
Defining
\begin{align}
 \tau_0 = \frac{\Tau_0}{M} = \frac{k_T}{M} e^{-|\eta_k - Y|}
\,,\end{align}
the phase space bound from \eq{x12_FKS} can be written as
\begin{align}
 x_1 &= \bar x_1 \left( \tau_0 e^{|\eta_k - Y|} \, e^{+ (\eta_k - Y)} + \sqrt{1+ \tau_0^2 e^{2|\eta_k - Y|}} \right) \le 1
\,,\nn\\
 x_2 &= \bar x_2 \left( \tau_0 e^{|\eta_k - Y|} \, e^{- (\eta_k - Y)} + \sqrt{1+ \tau_0^2 e^{2|\eta_k - Y|}} \right) \le 1
\,.\end{align}
By using $Y = \frac12 \ln(\bar x_1 / \bar x_2)$, we can write
$\bar x_{1,2} = \sqrt{\bar x_1 \bar x_2} e^{\pm Y}$. Thus, we obtain the symmetrical result
\begin{align} \label{eq:tau0_constraint_1}
 \tau_0 e^{|\eta_k - Y|} \, e^{\pm(\eta_k - Y)} + \sqrt{1+ \tau_0^2 e^{2|\eta_k - Y|}}
 &\le \frac{e^{\mp Y}}{\sqrt{\bar x_1 \bar x_2}}
\,.\end{align}
Defining $s \equiv \sgn(\eta_k - Y)$, the two cases $s = \pm 1$ can be combined as
\begin{align} \label{eq:tau0_constraint_2}
 \tau_0 e^{2|\eta_k - Y|} + \sqrt{1+ \tau_0^2 e^{2|\eta_k - Y|}}
 &\le \frac{e^{- s Y}}{\sqrt{\bar x_1 \bar x_2}}
   \equiv c(+s)
\,,\nn\\
 \tau_0 + \sqrt{1+ \tau_0^2 e^{2|\eta_k - Y|}}
 &\le \frac{e^{s Y}}{\sqrt{\bar x_1 \bar x_2}}
   \equiv c(-s)
\,.\end{align}
For brevity, we defined $c(s) = \bigl(\sqrt{\bar x_1 \bar x_2} e^{s Y}\bigr)^{-1}$ such that
\begin{align}
 c(+1) = \bigl(\sqrt{\bar x_1 \bar x_2} e^{ Y}\bigr)^{-1} = \frac{1}{\bar x_1}
 \,,\qquad
 c(-1) = \bigl(\sqrt{\bar x_1 \bar x_2} e^{- Y}\bigr)^{-1} = \frac{1}{\bar x_2}
\,.\end{align}
Since the constraints in \eq{tau0_constraint_2} are monotonically increasing with $|\eta_k - Y|$,
we can solve them fairly easily. We obtain
\begin{align} \label{eq:tau0_constraint_3}
 |\eta_k - Y| &\le \Delta Y_{\rm max}[\sgn(\eta_k - Y)]
\,,\\\nn
 \Delta Y_{\rm max}(s) &
 = \min \left\{ \frac12 \ln \frac{\tau_0 + 2 c(s) - \sqrt{\tau_0^2 + 4 c(s) \tau_0 + 4}}{2 \tau_0}
                 \,,\,
                 \frac12 \ln \frac{[\tau_0 - c(-s)]^2 - 1}{\tau_0^2} \right\}
\,.\end{align}
For this bound to be positive, we have to impose that
\begin{align}
 c(s) \ge1 \qquad\land\qquad |\tau_0 - c(-s)| > 1
\,,\end{align}
where the first constraint is always fulfilled since $c(\pm1) = \frac{1}{\bar x_{1,2}} \ge 1$.
To construct the final solution, consider both cases separately:
\begin{align}
 &\eta_k > Y: \qquad \eta_k - Y \le +\Delta Y_{\rm max}(+1)
             \quad\Leftrightarrow\quad
             \eta_k \le Y + \Delta Y_{\rm max}(+1)
\,,\nn\\
 &\eta_k < Y: \qquad \eta_k - Y \ge -\Delta Y_{\rm max}(-1)
             \quad\Leftrightarrow\quad
             \eta_k \ge Y - \Delta Y_{\rm max}(-1)
\,.\end{align}
In summary, we obtain that
\begin{align}
 \eta_{\rm min} \le \eta_k \le \eta_{\rm max}
\,,\end{align}
with
\begin{align}
 \eta_{\rm min} = Y - \Delta Y_{\rm max}(-1)
\,,\qquad
 \eta_{\rm max} = Y + \Delta Y_{\rm max}(+1)
\,.\end{align}
To ensure that the solution is physical, we have to require that
\begin{align}
 \left|\tau_0 - \frac{1}{\bar x_{1,2}} \right| > 1
 \qquad\land\qquad \eta_{\rm min} \le \eta_{\rm max}
\,.\end{align}

For the actual numerical implementation we perform the variable transform
\begin{align}
 t = \tanh(\eta_k - Y)
\,,\end{align}
which maps $\eta_k \in (-\infty, \infty)$ into the finite range $t \in (-1,1)$.
Taking the Jacobian into account, we find
\begin{align} \label{eq:Fcorr_7}
 \frac{J_\ell(\Phi_{\rm FJ})}{F_\ell^{\rm corr}(\Phi_{\rm FJ})} &
 = \sum_{\ell'} \frac{\Tau_0}{8 \pi^2} \int_{t_{\rm min}}^{t_{\rm max}}
   \frac{\df t}{1-t^2} \frac{1+|t|}{1-|t|} \, J_{\ell '}(\bar\Phi'_{\rm FJ})
 \nn\\&
 = \sum_{\ell'} \frac{\Tau_0}{8 \pi^2} \int_{t_{\rm min}}^{t_{\rm max}}
   \frac{\df t}{(1-|t|)^2} \, J_{\ell '}(\bar\Phi'_{\rm FJ})
\,,\end{align}
and the integration bounds are given by
\begin{align}
 t_{\rm min} &= \tanh[-\Delta Y_{\rm max}(-1)] = \frac{T_{\rm min} - 1}{T_{\rm min} + 1}
\,,\nn\\
 t_{\rm max} &= \tanh[+\Delta Y_{\rm max}(+1)] = \frac{T_{\rm max} - 1}{T_{\rm max} + 1}
\,,\end{align}
where $\Delta_\pm \equiv \Delta Y_{\rm max}(\pm1)$ and
\begin{align} \label{eq:t_bounds}
 \frac{1}{T_{\rm min}} &
 = e^{2 \Delta Y_{\rm max}(-1)]}
 = \min \left\{ \frac{\tau_0 + 2 \bar x_2^{-1} - \sqrt{\tau_0^2 + 4 \bar x_2^{-1} \tau_0 + 4}}{2 \tau_0}
                 \,,\,
               \frac{[\tau_0 - \bar x_1^{-1}]^2 - 1}{\tau_0^2} \right\}
\,,\nn\\
 T_{\rm max} &
 = e^{2 \Delta Y_{\rm max}(+1)]}
 = \min \left\{ \frac{\tau_0 + 2 \bar x_1^{-1} - \sqrt{\tau_0^2 + 4 \bar x_1^{-1} \tau_0 + 4}}{2 \tau_0}
                 \,,\,
               \frac{[\tau_0 - \bar x_2^{-1}]^2 - 1}{\tau_0^2} \right\}
\,.\end{align}
The solution is physical if and only if
\begin{align}
 T_{\rm min} > 0
 \quad\land\quad
 T_{\rm max} > 0
 \quad\land\quad
 T_{\rm min} \le T_{\rm max}
\,,\end{align}
or equivalently
\begin{align}
 -1 < t_{\rm min} < -1
 \quad\land\quad
 -1 < t_{\rm max} < -1
 \quad\land\quad
 t_{\rm min} \le t_{\rm max}
\,.\end{align}
Finally, note that with this change of variables
\begin{align}
 \kappa = e^{2 (\eta_k - Y)} = \frac{1+t}{1-t}
\,,\end{align}
with which \eq{xi_eta} becomes
\begin{align} \label{eq:xi_eta_2}
 \xi = 1 - \frac{1}{1 + \beta \eps^2}
\,,\qquad
 y = t \left(1 - \frac{2}{\beta}\right)
\,,\qquad
 \beta = 2 + 2 t \sqrt{\frac{1 + \eps^2}{1-t^2}}
\,,\end{align}
with
\begin{align}
 \eps^2
 = \frac{k_T^2}{M^2}
 = \tau_0^2 e^{2|\eta_k - Y|}
 = \tau_0^2 \frac{1+|t|}{1-|t|}
\,.\end{align}

\subsection[Application to $p_T$]
              {\boldmath Application to \texorpdfstring{$p_T$}{pT}}

We finally note that we can obtain a simpler result
than the one in appendix A of \cite{Monni:2019whf} starting from the relation
\begin{align} \label{eq:Fcorr_pT_1}
 \left[\frac{J_\ell(\Phi_{\rm FJ})}{F_\ell^{\rm corr}(\Phi_{\rm FJ})}\right]_{p_T} &
 \stackrel{!}{=} e^{-|Y - \eta_k|} \left[\frac{J_\ell(\Phi_{\rm FJ})}{F_\ell^{\rm corr}(\Phi_{\rm FJ})}\right]_{\Tau_0 = p_T e^{-|Y - \eta_k|}}
 \nn\\&
 = \sum_{\ell'} \frac{p_T}{8 \pi^2} \int\!\df \eta_k \, J_{\ell '}(\bar\Phi'_{\rm FJ})
   \Theta\left( M \eps_{\rm max} - p_T \right)
\,.\end{align}
The $\Theta$ constraint is different from the one in \eq{Fcorr_5}
due to the missing factor $e^{|Y-\eta_k|}$. 
We can actually solve this bound analytically.
Using \eq{eps_max}, the constraint reads
\begin{align}
 \eps \le \frac{\sqrt{1 - (1-\kappa) \bar x_1^2} - \sqrt{\kappa}}{\bar x_1 (1-\kappa)}
 \quad\land\quad
 \eps \le \frac{\sqrt{1 - (1-\kappa^{-1}) \bar x_2^2} - \sqrt{\kappa^{-1}}}{\bar x_2 (1-\kappa^{-1})}
\,.\end{align}
Where here $\kappa = e^{2 \eta_k} \bar x_2 / \bar x_1$ and $\eps = p_T/M$.
Both constraints are monotonic with $\kappa$, so solving for equality gives the bounds on $\eta_k$.
The quadratic equations are solved easily and yield
\begin{align} \label{eq:eta_min_max}
 \eta_{\rm min} = -\ln\frac{1- \bar x_2 \sqrt{1 + \eps^2}}{\eps \sqrt{\bar x_1 \bar x_2}}
\,,\qquad
 \eta_{\rm max} = \ln\frac{1- \bar x_1 \sqrt{1 + \eps^2}}{\eps \sqrt{\bar x_1 \bar x_2}}
\,.\end{align}
These solutions are only physical if
\begin{align}
 \eps \le \frac{\sqrt{1-\bar x_{1,2}}}{\bar x_{1,2}}
 \qquad \land \qquad
 \eta_{\rm min} < \eta_{\rm max}
\,.\end{align}
With these constraints, \eq{Fcorr_pT_1} becomes
\begin{align} \label{eq:Fcorr_pT_2}
 \left[\frac{J_\ell(\Phi_{\rm FJ})}{F_\ell^{\rm corr}(\Phi_{\rm FJ})}\right]_{p_T} &
 = \sum_{\ell'} \frac{p_T}{8 \pi^2} \int_{\eta_{\rm min}}^{\eta_{\rm max}} \df \eta_k \, J_{\ell '}(\bar\Phi'_{\rm FJ})
\,.\end{align}
In practice, it is useful to restrict to perform a variable transformation
in \eq{Fcorr_pT_3} to have a finite integration range. We choose the transform
\begin{align}
 t : (-\infty,\infty) \to (-1,1)
 \,,\quad
 \eta \mapsto t(\eta) = \tanh(\eta)
\,.\end{align}
Taking the Jacobian into account, \eq{Fcorr_pT_2} becomes
\begin{align} \label{eq:Fcorr_pT_3}
 \left[\frac{J_\ell(\Phi_{\rm FJ})}{F_\ell^{\rm corr}(\Phi_{\rm FJ})}\right]_{p_T} &
 = \sum_{\ell'} \frac{p_T}{8 \pi^2} \int_{t_{\rm min}}^{t_{\rm max}} \frac{\df t}{1-t^2} \, J_{\ell '}(\bar\Phi'_{\rm FJ})
\,,\end{align}
with the integration bounds given by
\begin{align} \label{eq:t_min_max}
 t_{\rm min} &= \tanh\eta_{\rm min} = \frac{e^{2 \eta_{\rm min}} - 1}{e^{2\eta_{\rm min}} + 1}
\,,\nn\\
 t_{\rm max} &= \tanh\eta_{\rm max} = \frac{e^{2 \eta_{\rm max}} - 1}{e^{2\eta_{\rm max}} + 1}
\,.\end{align}

\section{\minnlo{} for colour-singlet plus jet production using $1$-jettiness}
\label{sec:higgsplusjet}

Here, we briefly review $1$-jettiness factorization and the associated evolution equations,
and derive the \minnlo~method based on $\Tau_1$. The steps proceed analogously
to the case of $\Tau_0$ that was presented in detail in \sec{H0jtau0},
to which we refer for more details on key steps of the derivation.

\subsection{Review of $\Tau_1$ factorization and evolution}
\label{sec:tau1_review}

We begin by reviewing the resummation formalism for 
$1$-jettiness $\Tau_1$~\cite{Stewart:2010tn}.
We denote the underlying $1$-jet Born process and corresponding momenta as
\begin{align} \label{eq:1jet_Born}
 a(q_a) + b(q_b) \to F(q) + j(q_j)
\,,\end{align}
where $a$ and $b$ label the incoming partons with momenta $q_a$ and $q_b$,
respectively, $j$ is the parton initiating the outgoing jet with momentum $q_j$,
and $F$ is the colour-singlet final state of total momentum $q$.
For an event with $M \ge 1$ final-state partons, one requires an infrared-safe
projection onto the $1$-jet configuration in \eq{1jet_Born}.
The projection determines a $1$-jet kinematics of the form in \eq{1jet_Born}
with massless reference vectors $q_i^\mu = E_i (1, \vec n_i)$,%
\footnote{If the jet clustering yields massive jets, one can trivially construct
massless reference vectors as $q_i^\mu = E_i n_i^\mu$ with $n_i^\mu = (1, \vec P_i / |\vec P_i|)$,
where $E_i$ and $\vec P_i$ are the energy and momentum of the jet, respectively.}
where the incoming momenta $q_{a,b}$ are always aligned along the beam axes.
Following \citere{Jouttenus:2011wh}, we define $\Tau_1$ as
\begin{align} \label{eq:def_Tau1}
 \Tau_1
 = \sum_k \min_{i\in\{a,b,j\}} \Bigl\{ \frac{2 q_i \cdot p_k}{Q_i} \Bigr\}
 = \sum_k \min\biggl\{ \frac{2 q_a \cdot p_k}{Q_a}, \frac{2 q_b \cdot p_k}{Q_b}, \frac{2 q_j \cdot p_k}{Q_j} \biggr\}
\,,\end{align}
where the sum runs over all final-state hadronic momenta $p_k$,
and $q_i$ are the Born-like reference momenta as described above.
The normalization factors $Q_i$ in \eq{def_Tau1} allow for different definitions of $1$-jettiness.
Different choices of the $Q_i$ and algorithms for determining the $q_i$
only affect power corrections to the factorization theorem, but not its functional form.
For later convenience, we also define normalized directions and their scalar products as
\begin{align} \label{eq:q_hat}
 \hat q_i^\mu = \frac{q_i^\mu}{Q_i}
\,,\qquad
 \hat s_{ij} = 2 \hat q_i \cdot \hat q_j = \frac{2 q_i \cdot q_j}{Q_i Q_j}
\,.\end{align}
Note that all $q_i$ correspond to physical momenta, and hence all $q_i \cdot q_j > 0$.

$1$-jettiness obeys a factorization theorem in the limit $\Tau_1 \to 0$~\cite{Stewart:2010tn},
\begin{align} \label{eq:Tau1_fact_1}
 \frac{\df\sigma^\sing}{\df\Phi_{\rm FJ} \df \Tau_1}
 = \sum_\kappa \frac{\df|\cM_\kappa|^2}{\df \Phi_{\rm FJ}} H_\kappa(\Phi_{\rm FJ}, \mu)
   \int\df t_a \df t_b \df t_j \, & B_a(t_a, x_a, \mu) B_b(t_b, x_b, \mu)  J_j(t_j, \mu)
 \nn\\&\times
   S_\kappa\Bigl(\Tau_1 - \sum_i \frac{t_i}{Q_i} , \{\hat q_i\}, \mu\Bigr)
\,.\end{align}
Here, the sum runs over all flavour structures $\kappa \equiv \{a, b, j\}$
contributing to the process in \eq{1jet_Born}, and $\cM_\kappa$ and $H_\kappa$
denote the corresponding matrix element and hard function, respectively.
In \eq{Tau1_fact_1}, $B_{a,b}$ are the same beam functions appearing for $\Tau_0$,
$J_j$ is the jet function, and $S_\kappa$ is the $\Tau_1$ soft function.
Note that the jet function only differs between quarks and gluons,
while the soft function is sensitive to the full flavour structure $\kappa$ of the process,
as well as the Born reference momenta $\{\hat q_i\}$ and the normalization factors $Q_i$.
Performing the same Fourier transform as in \eq{FT},
\begin{align} \label{eq:FT_Tau1}
 B_i(y, x, \mu) &
 = \int\df t \, e^{-\img t y} B_i(t, x, \mu)
\,,\nn\\
 J_j(y,  \mu) &
 = \int\df t \, e^{-\img t y} J_j(t, \mu)
\,,\nn\\
 S_\kappa(y,  \{\hat q_i\}, \mu) &
 = \int\df \Tau \, e^{-\img \Tau y} S_\kappa(\Tau, \{\hat q_i\}, \mu)
\,,\end{align}
\eq{Tau1_fact_1} can be written as
\begin{align} \label{eq:Tau1_fact_2}
 \frac{\df\sigma^\sing}{\df\Phi_{\rm FJ} \df \Tau_1}
 = \sum_\kappa \frac{\df|\cM_\kappa|^2}{\df \Phi_{\rm FJ}} H_\kappa(\Phi_{\rm FJ}, \mu)
   \int\frac{\df y}{2\pi} e^{\img y \Tau_1} \, &
   B_a\Bigl(\frac{y}{Q_a},x_a,\mu\Bigr) B_b\Bigl(\frac{y}{Q_b},x_b,\mu\Bigr) J_j\Bigr(\frac{y}{Q_j}, \mu\Bigr)
 \nn\\&\times
   S_\kappa(y, \{\hat q_i\}, \mu)
\,.\end{align}

In Fourier space, the RGEs of the hard, beam, jet and soft functions are given by
\begin{alignat}{2} \label{eq:Tau1_RGEs}
 \frac{\df}{\df\ln\mu} \ln H_\kappa(\Phi_{\rm FJ}, \mu)
 & = \gamma_H^\kappa(\{q_i\}, \mu)
\,,\qquad~
 \frac{\df}{\df\ln\mu} S_\kappa(y, \{\hat q_i\}, \mu)
 = \gamma_S^\kappa(y, \{\hat q_i\}, \mu)
\,,\nn\\
 \frac{\df}{\df\ln\mu} \ln B_i\Bigl(\frac{y}{Q_i},x,\mu\Bigr)
 &= \gamma_B^i\Bigl(\frac{y}{Q_i},\mu\Bigr)
\,,\qquad
 \frac{\df}{\df\ln\mu} \ln J_j\Bigl(\frac{y}{Q_j},\mu\Bigr)
   = \gamma_J^j\Bigl(\frac{y}{Q_j},\mu\Bigr)
\,.\end{alignat}
In \eqn{eq:Tau1_RGEs} we have made explicit the flavour index $i$, $j$ for initial and final state legs for clarity.
We write the anomalous dimensions as~\cite{Becher:2019avh}
\begin{align} \label{eq:Tau1_anom_dims}
 \gamma_H^\kappa(\{q_i\}, \mu) &
 = (n_q C_F + n_g C_A) \hat \Gamma_C[\as(\mu)] \ln\frac{Q^2}{\mu^2}
   + \gamma_H^\kappa[\{q_i\}, \as(\mu)]
\,,\nn\\
 \gamma_B^i\Bigl(\frac{y}{Q_i},\mu\Bigr) &
 = 2 \GammaC^i[\as(\mu)] \ln\frac{y \mu^2}{Q y_0} + \gamma_B^i[Q_i, \as(\mu)]
 \nn\\&
 = \gamma_J^i\Bigl(\frac{y}{Q_i},\mu\Bigr)
\,,\nn\\
 \gamma_S^\kappa(y, \{\hat q_i\}, \mu) &
 = -2 (n_q C_F + n_g C_A) \hat \Gamma_C[\as(\mu)] \ln\frac{y \mu}{y_0}
   + \gamma_S^\kappa[\{\hat q_i\}, \as(\mu)]
\,.\end{align}
Here, as before $y_0 = -\img e^{-\gamma_E}$, $n_q$ and $n_g$ denote the number of quarks
and gluons for the flavour structure of $\kappa$, respectively, and $\hat \Gamma_C$ is
the colour-stripped cusp anomalous dimension. The noncusp anomalous dimensions read
\begin{align} \label{eq:Tau1_anom_dim_noncusp}
 \gamma_H^\kappa(\{q_i\}, \as) &
 = - \hat \Gamma_C(\as) \sum_{i \ne k} (\bT_i \cdot \bT_k) \ln\frac{2 q_i \cdot q_k}{Q^2}
   + \gamma_H^\kappa(\as)
\,,\nn\\
 \gamma_B^i(Q_i, \as) &
 = \GammaC^i[\as(\mu)] \ln\frac{Q^2}{Q_i^2} +  \gamma_B^i(\as)
\,,\nn\\
 \gamma_S^\kappa(\{\hat q_i\}, \as)&
 = \hat \Gamma_C[\as(\mu)] \sum_{i \ne k} (\bT_i \cdot \bT_k) \ln\frac{2 q_i \cdot q_k}{Q_i Q_k}
   + \gamma_S^\kappa(\as)
\,.\end{align}
The $\bT_i$ are the colour-charge operators,
and the appearing products can be evaluated using
\begin{align} \label{eq:TiTj}
 \bT_i \cdot \bT_k = \frac12 (\bT_l^2 - \bT_i^2 - \bT_k^2) \,,\qquad  l \ne i, k
\,,\end{align}
which follows from colour conservation of the $2\to1$ process.
Here, $\gamma_H^\kappa(\as), \gamma_B^i(\as)$ and $\gamma_S^\kappa(\as)$
are the standard noncusp anomalous dimensions, while the logarithmic terms
encode the measure dependence. The hard and soft anomalous dimensions are reported in Appendix~\ref{app:constants}, whereas the noncusp anomalous dimension of the beam  function, which is identical to that of the jet function, read, up to two loops~\cite{Fleming:2003gt,Becher:2006mr,Becher:2009th,Stewart:2010qs,Berger:2010xi}
\begin{align} \label{eq:gamma_J_coeffs}
 \gamma_B^{q\,(1)} &
 = 3 C_F
\,,\\\nn
 \gamma_B^{q\,(2)} &
 = C_F \left[ C_A  \left(-20 \zeta_3+\frac{11 \pi ^2}{18}+\frac{1769}{108}\right)
             + C_F \left(12 \zeta_3-\pi ^2+\frac{3}{4}\right) - n_f\frac{121 + 6 \pi^2}{54} \right]\,,\\
 \gamma_B^{g\,(1)} &
 = 4 \pi \beta_0\,,\\\nn
 \gamma_B^{g\,(2)} &
 = C_A \left[ C_A \left(-8 \zeta_3-\frac{11 \pi ^2}{18}+\frac{548}{27}\right)
             + \frac{n_f}{27} \left(3 \pi ^2-92\right) \right]
 - C_F n_f
\,.\end{align}

To minimize the logarithms in \eq{Tau1_anom_dims}, we choose the same
canonical resummation scales as in \eq{canonical_scales_Tau0},
\begin{align} \label{eq:canonical_scales_Tau1}
 \mu_H = Q
\,,\qquad
 \mu_{B_a} = \mu_{B_b} = \mu_J = \sqrt{\frac{Q y_0}{y}} = \sqrt{\mu_H \mu_S}
\,,\qquad
 \mu_S = \frac{y_0}{y}
\,.\end{align}
Furthermore, we also evolve the hard and soft functions such that they are
evaluated in terms of $\as(\mu_B)$. To be precise, we first define the hard, jet
and soft functions at their respective canonical scales as
\begin{alignat}{2} \label{eq:H_S_canonical_Tau1}
 \bar H_\kappa(\Phi_{\rm FJ}, Q)
 &\equiv H_\kappa(\Phi_{\rm FJ}, \mu_H = Q)
 &&= \sum_{n=0}^\infty \left[\frac{\as(Q)}{2\pi}\right]^n H_\kappa^{(n)}(\Phi_{\rm FJ})
\,,\nn\\
 \bar J_j\Bigl(\frac{Q}{Q_j}, \mu_B \Bigr)
 &\equiv J_j\Bigl(\frac{y}{Q_j}, \mu = \mu_B \Bigr)
 &&= \sum_{n=0}^\infty \left[\frac{\as(\mu_J)}{2\pi}\right]^n J_j^{(n)}(Q/Q_j)
\,,\nn\\
 \bar S_\kappa(\{\hat q_i\}, y_0/y)
 &\equiv S_\kappa(y, \{\hat q_i\}, \mu_S = y_0/y)
 &&= \sum_{n=0}^\infty \left[\frac{\as(y_0/y)}{2\pi}\right]^n S_\kappa^{(n)}(\{\hat q_i\})
\,,\end{alignat}
where $\Phi_{\rm FJ}$ encodes the dependence on the kinematics of the Born $F$+jet configuration.
For clarity, we use the modified symbols $\bar H_\kappa$ and $\bar S_\kappa$ to distinguish
these functions from the original hard and soft functions $H_\kappa$ and $S_\kappa$.
Importantly, the right-hand side depends on $Q$ and $y$ only through $\as(Q)$ and $\as(y_0/y)$, respectively.%
\footnote{Strictly speaking, $H_\kappa^{(n)}(\Phi_{\rm FJ})$ also depends on $Q$ through $\Phi_{\rm FJ}$.}
We can now evolve these functions to the beam scale similar to \eq{shifted_H_S},
\begin{align} \label{eq:shifted_H_S_Tau1}
 \bar H_\kappa(\Phi_{\rm FJ}, Q) &
 = \bar H_\kappa(\Phi_{\rm FJ}, \mu_B) \exp\left[ \int_{\mu_B}^{Q} \frac{\df\mu'}{\mu'} \gamma_{\bar H_\kappa}[\Phi_{\rm FJ}, \as(\mu')] \right]
\,,\nn\\
 \bar S_\kappa(\{\hat q_i\}, y_0/y) &
 = \bar S_\kappa(\{\hat q_i\}, \mu_B) \exp\left[ \int_{\mu_B}^{y_0/y} \frac{\df\mu'}{\mu'} \gamma_{\bar S_\kappa}[\{\hat q_i\}, \as(\mu')] \right]
\,,\end{align}
where the anomalous dimensions are given by
\begin{alignat}{2}
 \gamma_{\bar H_\kappa}[\Phi_{\rm FJ}, \as(\mu)] &
 = \frac{\df\ln \bar H_\kappa(\Phi_{\rm FJ}, \mu)}{\df\ln\mu}
 &&= -4 \pi \beta_0 \frac{H_\kappa^{(1)}(\Phi_{\rm FJ})}{H_\kappa^{(0)}(\Phi_{\rm FJ})} \left[\frac{\as(\mu)}{2\pi}\right]^2 + \cO(\as^3)
\,,\nn\\
 \gamma_{\bar S_\kappa}[\{\hat q_i\}, \as(\mu)] &
 = \frac{\df\ln \bar S_\kappa(\{\hat q_i\}, \mu)}{\df\ln\mu}
 &&= -4 \pi \beta_0 \frac{S_\kappa^{(1)}(\{\hat q_i\})}{S_\kappa^{(0)}(\{\hat q_i\})} \left[\frac{\as(\mu)}{2\pi}\right]^2 + \cO(\as^3)
\,.\end{alignat}
By making the choice in \eq{canonical_scales_Tau1} and evolving all functions
to the common scale $\mu = \mu_B$, we obtain
\begin{align} \label{eq:Tau1_resummed_2}
 \frac{\df\sigma^\sing}{\df\Phi_{\rm FJ} \, \df\Tau_1} &
 = \sum_\kappa \int\frac{\df y}{2\pi} \, e^{\img y \Tau_1} \, \cL_\kappa(y_0/y) \, e^{-\cS_\kappa(y_0/y)}
\,.\end{align}
Here, the canonical luminosity is defined as
\begin{align} \label{eq:Tau1_L}
 \cL_\kappa(y_0/y) &
 = \frac{\df|\cM_\kappa|^2}{\df \Phi_{\rm FJ}} \bar H_\kappa(\Phi_{\rm FJ}, \mu_B)
   B_a\Bigl(\frac{y}{Q_a},x_a,\mu_B\Bigr) \, B_b\Bigl(\frac{y}{Q_b},x_b,\mu_B\Bigr) \,
   \bar J_j\Bigl(\frac{Q}{Q_j}, \mu_B \Bigr)\bar S_\kappa(\{\hat q_i\}, \mu_B)
 \nn\\&
 = \sum_{a', b'} \frac{\df|\cM_\kappa|^2}{\df \Phi_{\rm FJ}} \bar H_\kappa(\Phi_{\rm FJ}, \mu_B)
   (C \otimes f)_a \Bigl(\frac{y}{Q_a},x_a,\mu_B\Bigr)
   (C \otimes f)_b \Bigl(\frac{y}{Q_b},x_b,\mu_B\Bigr)
   \nn\\&\qquad\times
   \bar J_j\Bigl(\frac{Q}{Q_j}, \mu_B \Bigr) \frac{\bar S_\kappa(\{\hat q_i\}, \mu_B)}{S(\mu_B)}
\,.\end{align}
In the second step, we divided the $\Tau_1$ soft function $\bar S_\kappa$
by the $\Tau_0$ soft function $S$ as defined in \eq{H_S_canonical},
such that the $C_{ij}$ are the same matching coefficients of the beam functions onto the PDFs
as in \eq{beam_matching_2}. The canonical Sudakov form factor is defined as
\begin{align} \label{eq:Tau1_S}
 \cS_\kappa(y_0/y) &
 = 2 \int_{\sqrt{Q y_0/y}}^{Q} \frac{\df\mu'}{\mu'} \left\{
   A_\kappa[\as(\mu')] \ln\frac{Q^2}{\mu'^2} + B_H^\kappa[\Phi_{\rm FJ}, \as(\mu')]
   \right\}
 \nn\\&
   + 2 \int_{\sqrt{Q y_0/y}}^{y_0/y} \frac{\df\mu'}{\mu'} \left\{
      A_\kappa[\as(\mu')] \ln\frac{(y_0/y)^2}{\mu^2} + B_S^\kappa[\{\hat q_i\}, \as(\mu')]
   \right\}
\,.\end{align}
Here, the first exponential is the hard evolution, while the second is the soft evolution.
The $A$ and $B$ coefficients are given by
\begin{align} \label{eq:resum_coeffs_Tau1}
 A_\kappa(\as) &
 = \frac12 (n_q C_F + n_g C_A) \Gamma_C(\as)\,,\nn\\
 B_H^\kappa(\Phi_{\rm FJ}, \as) &
 = \frac12 \gamma_H^\kappa(\{q_i\}, \as)
   - \frac12 \gamma_{\bar H_\kappa}(\Phi_{\rm FJ}, \as)
\,,\nn\\
 B_S^\kappa(\{\hat q_i\}, \as) &
 = \frac12 \gamma_S^\kappa(\{\hat q_i\}, \as)
   - \frac12 \gamma_{\bar S_\kappa}(\{q_i\}, \as)
\,.\end{align}
Comparing \eqs{Tau1_L}{Tau1_S} to \eq{Tau0_L_S}, we see that the overall structure
of the resummed $\Tau_1$ spectrum is very similar to that of $\Tau_0$.
The key difference is the explicit dependence of the hard and soft functions and
their anomalous dimensions on the flavour channel $\kappa$, as well as the additional
jet function.

We complete this section by presenting explicit expressions for the jet and soft functions for $\mathcal T_1$. 

\paragraph{$\Tau_1$ jet function.}
The jet function RGE in \eq{Tau1_anom_dims} predicts that
\begin{align} \label{eq:jet_fo}
 J_i\Bigl(\frac{y}{Q_j}, \mu\Bigr) &
 = 1
 + \frac{\as(\mu)}{2\pi} \biggl[ \frac{L_j^2}{2} \Gamma_C^{i\,(1)} + \frac{L_j}{2} \gamma_{j}^{i\,(1)} + j_1^i\biggr]
 \nn\\&\quad
 + \biggl(\frac{\as(\mu)}{2\pi}\biggr)^2 \biggl\{
   \frac{L_j^4}{8} (\Gamma_C^{i\,(1)})^2
   + \frac{L_j^3}{4} \Gamma_C^{i\,(1)} \biggl(  \gamma_j^{i\,(1)} + \frac{4}{3} \pi \beta_0 \biggr)
   \nn\\&\hspace{2.5cm}
   + \frac{L_j^2}{2} \biggl[ \GammaC^{i\,(2)} + \GammaC^{i\,(1)} j_1^i + \frac{1}{4} \gamma_j^{i\,(1)} (\gamma_j^{i\,(1)} + 4 \pi \beta_0) \biggr]
   \nn\\&\hspace{2.5cm}
   + \frac{L_j}{2} \bigl[ \gamma_j^{i\,(2)} + j_1^i \bigl(\gamma_j^{i\,(1)} + 4 \pi \beta_0\bigr) \bigr]
   + j_2^i
   \biggr\}
 + \cO(\as^3)
\,.\end{align}
Here, we define $L_j = \ln[y \mu^2 / (Q_j y_0)]$, and $i=q,g$ distinguishes
between quarks and gluons. The cusp anomalous dimension is given in \eq{A_coeffs},
the $\beta_n$ are given in \eq{beta_coeffs}, and the noncusp anomalous dimensions
are provided in \eq{gamma_J_coeffs}.
The jet-function constants $j_n^i$ are known up to three loops~\cite{Bruser:2018rad,Banerjee:2018ozf},
and are collected up to two loops in \citere{Gaunt:2015pea}.
They are commonly expressed in momentum space, where instead of the logarithms $L_j$
in \eq{jet_fo} one obtains plus distributions. The conversion between these induces
additional terms. Taking these into account they are given by
\begin{align}
 j_1^q &
 = C_F \Bigl(\frac{7}{2} - \frac{\pi^2}{3}\Bigr)\,,\nn\\
 j_2^q &
 = C_F \left[C_A \left(-\frac{9 \zeta_3}{2}-\frac{1}{720} \pi ^2 \left(775+37 \pi ^2\right)+\frac{53129}{2592}\right)+\frac{\left(234 \pi ^2-4057\right) n_f}{1296}\right]
 \nn\\&\quad
  + C_F^2 \left(-\frac{3 \zeta_3}{2}+\frac{61 \pi ^4}{360}+\frac{205}{32}-\frac{97 \pi ^2}{48}\right)
\end{align}
for the quark case, while for gluons they read
\begin{align}
 j_1^g &
 = \frac{1}{18} \left(67-6 \pi ^2\right) C_A-\frac{5 n_f}{9}\,,\nn\\
 j_2^g &
 = \frac{C_A^2}{1296} \left(-9504 \zeta_3 +153 \pi ^4-4344 \pi ^2+40430\right)
  + \frac{1}{108} C_A n_f ( 67 \pi ^2 - 72 \zeta_3 - 760 )
  \nn\\&\quad
  + C_F n_f \left(2 \zeta_3 -\frac{55}{24}\right) + \frac{n_f^2}{162} \left(50-3 \pi ^2\right)
\,.\end{align}
The coefficients $J^{(n)}(Q/Q_j)$ in \eq{H_S_canonical_Tau1} are obtained
from \eq{jet_fo} by setting $L_j = \ln(Q/Q_j)$.

\paragraph{$\Tau_1$ soft function.}
The fixed-order structure of the single-differential soft function can be found in \citere{Gaunt:2015pea},
where it is expanded as
\begin{align}
 S_\kappa(k, \{\hat q_i\}, \mu)
 = \sum_n \Bigl[\frac{\as(\mu)}{4\pi}\Bigr]^n
   \left[ S_{\kappa,-1}^{(n)}(\{\hat q_i\}) \delta(k)
          + \sum_{m=0}^{2n-1} S_{\kappa,m}^{(n)} \frac{1}{\mu} \cL_n\Bigl(\frac{k}{\mu}\Bigr) \right]
\,.\end{align}
Here, the $S_{\kappa,m}^{(n)}$ are those given in \citere{Gaunt:2015pea},
which uses different conventions than in this work. We require the Fourier-transformed
soft function at its natural scale $\mu = \mu_s = y_0/y$ as defined in \eq{H_S_canonical_Tau1},
\begin{align}
 \bar S_\kappa(\{\hat q_i\}, y_0/y) &
 \equiv S_\kappa(y, \{\hat q_i\}, \mu_S = y_0/y)
 \nn\\&
 = 1 + \frac{\as(\mu)}{2\pi} S_\kappa^{(1)}(\{\hat q_i\})
     + \Bigl[\frac{\as(\mu)}{2\pi}\Bigr]^2 S_\kappa^{(2)}(\{\hat q_i\})
     + \cO(\as^3)
\,.\end{align}
Evaluating the Fourier transform using \eq{FT} and adjusting to our conventions, we obtain
\begin{align}
 S_\kappa^{(1)}(\{\hat q_i\}) &
 = \frac12 S_{\kappa,-1}^{(1)}(\{\hat q_i\}) - \mathbf{C} \, \hat \Gamma^{(1)} \frac{\pi^2}{6}
\,,\nn\\
 S_\kappa^{(2)}(\{\hat q_i\}) &
 = \frac{7\pi^4}{360}  \bigl(\mathbf{C} \, \hat \Gamma_C^{(1)} \bigr)^2
 + \mathbf{C} \, \hat \Gamma_C^{(1)} \left[ - 2 \hat \Gamma_C^{(1)} \mathbf{L} \zeta_3 - \frac{8}{3} \pi  \beta_0 \zeta_3 -\frac{\pi^2}{12} S_{\kappa,-1}^{(1)}(\{\hat q_i\}) \right]
 -\frac{\pi^2}{6} \mathbf{C} \, \hat \Gamma_C^{(2)}
 \nn\\&\quad
  + \bigl(\mathbf{L}\, \hat \Gamma_C^{(1)}\bigr) \left[ \frac{1}{3} \pi ^3 \beta_0
  + \frac{\pi^2}{12}  \bigl(\mathbf{L}\,\hat \Gamma_C^{(1)}\bigr) \right]
  + \frac14 S_{\kappa,-1}^{(2)}(\{\hat q_i\})
\,.\end{align}
where $S_{\kappa,-1}^{(2)}(\{\hat q_i\}$ can extracted from \citere{Campbell:2017hsw,Bell:2023yso} (an independent calculation of the two-loop coefficients has been used in \citere{Boughezal:2015eha}) and
\begin{align}
 \mathbf{C} = \sum_{i \in \{a,b,j \}}  C_i
\,,\qquad
 \mathbf{L} = \sum_{i \ne k} (\bT_i \cdot \bT_k) \ln\frac{2 q_i \cdot q_k}{Q_i Q_k}
\,.\end{align}

\subsection{\minnlo{} formalism based on $\Tau_1$}
\label{sec:derivation_MiNNLO_Tau1}

Due to the similarity between the resummed $\Tau_0$ and $\Tau_1$ formulae,
the derivation of the \minnlo~method using $\Tau_0$, presented in detail
in \sec{derivation_MiNNLO_Tau0}, carries over to the case of $\Tau_1$ almost unchanged.
Thus, here we only briefly review the key steps, while referring for more details
to \sec{derivation_MiNNLO_Tau0}.

The starting point for the \minnlo{} method is the cumulant of \eq{Tau1_resummed_2},
\begin{align} \label{eq:Tau1Cut_resummed_1}
 \frac{\df \sigma^{\rm sing}(\TauCut)}{\df\Phi_{\rm FJ}} &
 = \sum_\kappa \int_0^{\TauCut} \df\Tau_1 \int\frac{\df y}{2\pi} \, e^{\img y \Tau_1} \, \cL_\kappa(y_0/y) \, e^{-\cS_\kappa(y_0/y)}
\,.\end{align}
We expand it around $y_0/y \sim \TauCut$, i.e.\ in $L_y = \ln(\TauCut y/y_0) \ll 1$.
By keeping only $S_\kappa(\TauCut)$ exponentiated and expanding all other terms
at NNLO accuracy, we obtain [cf.~\eq{TauCut_resummed_3}]
\begin{align} \label{eq:TauCut1_resummed_2}
 \frac{\df \sigma^{\rm sing}(\TauCut)}{\df\Phi_{\rm FJ}} &
 = \sum_\kappa e^{-\cS_\kappa(\TauCut)}  \Bigl[
   \cL_\kappa(\TauCut) \Bigl( 1 - \frac{\zeta_2}{2} [(\cS_\kappa')^2 - \cS_\kappa''] - \zeta_3 \cS_\kappa' \cS_\kappa'' + \frac{3 \zeta_4}{16} (\cS_\kappa'')^2 + \frac{\zeta_3}{3} \cS_\kappa''' \Bigr)
   \nn\\&
   + \cL_\kappa'(\TauCut) \bigl( \zeta_2 \cS_\kappa' + \zeta_3 \cS_\kappa'' \bigr)
   - \frac{\zeta_2}{2} \cL_\kappa''(\TauCut)
   + \cO(\as^3) \Bigr]
\,,\end{align}
where the derivatives are defined in \eq{def_fn}.
Next, we evaluate all derivatives in \eq{TauCut1_resummed_2} and exponentiate
all resulting logarithms, while the remaining terms are absorbed by redefining
the luminosity. This yields
\begin{align} \label{eq:TauCut1_resummed_3}
 \frac{\df \sigma^{\rm sing}(\TauCut)}{\df\Phi_{\rm FJ}} &
 = \sum_\kappa \tilde \cL_\kappa(\TauCut) e^{-\cS_\kappa(\TauCut)}
\,.\end{align}
Defining $\mu_B = \sqrt{Q \TauCut}$ and $L_\Tau = \frac12 \ln(Q/\TauCut)$,
the modified luminosity is defined similar to \eq{Tau0_cL_final} as
\begin{align} \label{eq:Tau1_cL_final}
 \tilde\cL_\kappa(\TauCut) &
 = \frac{\df |\cM_\kappa|^2}{\df \Phi_{\rm FJ}}
   \tilde H_\kappa(\Phi_{\rm FJ}, \mu_B) J_j\Bigl(\frac{y}{Q_j},\mu_B\Bigr) \frac{\bar S_\kappa(\{\hat q_i\}, \mu_B)}{S(\mu_B)}\nn\\&\hspace{2cm}
   \Bigl[ (\tilde C \otimes f)_a(x_a, \mu_B) \, (\tilde C \otimes f)_b(x_b, \mu_B)
   \nn\\&\hspace{2cm}
         -\zeta_2 (\hat P \otimes f)_a(x_a, \mu_B) \, (\hat P \otimes f)_b(x_b, \mu_B) \Bigr]
 \nn\\&
 - \frac{\df |\cM_\kappa|^2}{\df \Phi_{\rm FJ}}\Bigl(\frac{\as}{2\pi}\Bigr)^2 c^{'\kappa}_{1,1} L_\Tau
   \Bigl[ \bigl(\hat P^{(0)} \otimes f)_a(x_a, \mu_B) \, f_{b}(x_b, \mu_B)\nn\\&\hspace{2cm}
          + f_{a}(x_a, \mu_B) (\hat P^{(0)} \otimes f)_b(x_b, \mu_B) \Bigr]
\,,\end{align}
where the modified hard function and matching coefficients follow from \eq{def_tildeH},
\begin{align} \label{eq:def_tildeH_Tau1}
 \tilde H_\kappa(\Phi_{\rm FJ}, \mu_B) &
 = \bar H_\kappa(\Phi_{\rm FJ}, \mu_B) \biggl[ 1 + \frac{\as}{2\pi} c^\kappa_{1,0} + \left(\frac{\as}{2\pi}\right)^2 c^\kappa_{2,0} + \cO(\as^3) \biggr]
\,,\\ \nn
 \tilde C_{ij}(x, \mu_B) &
 = C_{ij}(x, \mu_B)
   - \Bigl(\frac{\as}{2\pi}\Bigr)^2 \left[
      \frac{\zeta_2}{2} \bigl(\hat P^{(0)} \otimes \hat P^{(0)}\bigr)_{ij}(x)
      + \bigl( c^{'\kappa}_{1,0} - \zeta_2 \pi \beta_0 \bigr)\hat P^{(0)}_{ij}(x)
   \right]
 + \cO(\as^3)
\,.\end{align}
The Sudakov factor is given by
\begin{align} \label{eq:Tau1_cS_final}
 \cS(\TauCut) &
 = 2 \int_{\sqrt{Q \TauCut}}^Q \frac{\df\mu'}{\mu'} \left[ A_\kappa[\as(\mu')] \ln\frac{Q^2}{\mu'^2} +B_H^\kappa[\Phi_{\rm FJ}, \as(\mu')] \right]
 \nn\\&
 + 2 \int_{\sqrt{Q \TauCut}}^{\TauCut} \, \frac{\df\mu'}{\mu'} \left[ A_\kappa[\as(\mu')] \ln\frac{(\TauCut)^2}{\mu'^2} + B_S^\kappa[\{\hat q_i\}, \as(\mu')] \right]
\,,\end{align}
where the $A_\kappa$ coefficients are given in \eq{resum_coeffs_Tau1} and  the $B_{H,S}^\kappa$ are defined in \eq{resum_coeffs_Tau1}.
The constants $c^\kappa_{n,m}$ and $c^{'\kappa}_{n,m}$ appearing in eqs.~\eqref{eq:Tau1_cL_final} -- \eqref{eq:def_tildeH_Tau1}
are identical to those in \eq{Tau0_constants}, up to changing the $A$ and $B$ coefficients
to those in \eq{resum_coeffs_Tau1}. In particular, they inherit the dependence of the $A_\kappa$
and $B^\kappa_{H,S}$ coefficients on the flavour channel $\kappa$ and the phase space $\Phi_{\rm FJ}$, which is kept implicit.

\addcontentsline{toc}{section}{References}
\bibliography{MiNNLO}
\bibliographystyle{JHEP}

\end{document}